\documentclass{aa}  

\usepackage{graphicx}
\usepackage{natbib}
\usepackage{txfonts}
\usepackage{placeins}
\bibpunct{(}{)}{;}{a}{}{,} 

\begin{document}

   \title{Probing the physics of  narrow-line regions of Seyfert galaxies I: The case of NGC~5427}

 \author{
  Michael A. Dopita\inst{1,2}
  \and Julia Scharw\"achter\inst{3}
  \and Prajval Shastri\inst{4}
  \and Lisa J. Kewley\inst{1,5}
  \and Rebecca Davies\inst{1}
  \and Ralph Sutherland\inst{1}
  \and  Preeti Kharb\inst{4}
  \and Jessy Jose\inst{4}			
  \and Elise Hampton\inst{1}
  \and Chichuan Jin\inst{6}
  \and Julie Banfield\inst{7}
  \and Hassan Basurah\inst{2}
  \and Sebastian Fischer\inst{8}
  }

\offprints{ \email{ Michael.Dopita@anu.edu.au}}

\institute{RSAA, The Australian National University, Cotter Road., Weston Creek, ACT 2611, Australia
  \and Astronomy Department, King Abdulaziz University, P.O. Box 80203, Jeddah, Saudi Arabia
  \and LERMA, Observatoire de Paris, 61 Avenue de l'Observatoire, 75014 Paris, France
  \and Indian Institute of Astrophysics, Koramangala 2B Block, Madiwala, Bangalore, Karnataka 560034, India
  \and Institute for Astronomy, University of Hawaii, 2680 Woodlawn Drive, Honolulu, HI 96822, USA
  \and Department of Physics, University of Durham, South Road, Durham DH1 3LE, United Kingdom
  \and CSIRO Astronomy \& Space Science, P.O. Box 76, Epping NSW, 1710 Australia
  \and German Aerospace Center (DLR), K\"onigswinterer Str. 522-524, 53227 Bonn
  }

   \date{}

 
  \abstract
   {The spectra of the extended narrow-line regions (ENLRs) of Seyfert 2 galaxies probe the physics of the central active galaxy nucleus (AGN), since they encode the energy distribution of the ionising photons, the radiative flux and radiation pressure, nuclear chemical abundances and the mechanical energy input of the (unseen) central AGN.}
   {We aim to constrain the chemical abundance in the interstellar medium of the ENLR by measuring the abundance gradient in the circum-nuclear \ion{H}{ii} regions to determine the nuclear chemical abundances, and to use these to in turn determine the EUV spectral energy distribution for comparison with theoretical models.}
   {We have used the Wide Field Spectrograph (WiFeS) on the ANU 2.3m telescope at Siding Spring to observe the nearby, nearly face-on, Seyfert 2 galaxy, NGC~5427. We have obtained integral field spectroscopy of both the nuclear regions and the \ion{H}{ii} regions in the spiral arms. The observed spectra have been modelled using the MAPPINGS IV photoionisation code, both to derive the chemical abundances in the \ion{H}{ii} regions and the Seyfert nucleus, and to constrain the EUV spectral energy distribution of the AGN illuminating the ENLR.}
   {We find a very high nuclear abundance, 3.0 times solar, with clear evidence of a nuclear enhancement of N and He, possibly caused by massive star formation in the extended ($\sim 100$pc) central disk structure. The circum-nuclear narrow-line region spectrum is fit by a radiation pressure dominated photoionisation model model with an input EUV spectrum from a Black Hole with mass $5\times10^7 M_{\odot}$ radiating at $\sim 0.1$ of its Eddington luminosity. The bolometric luminosity is closely constrained to be $\log L_{\mathrm bol.} = 44.3\pm 0.1$ erg~s$^{-1}$. The EUV spectrum characterised by a soft accretion disk and a harder component extending to above 15keV. The ENLR region is extended in the NW-SE direction. The line ratio variation in circum-nuclear spaxels can be understood as the result of mixing \ion{H}{ii} regions with an ENLR having a radius-invariant spectrum.}
   {}


   \maketitle
%

\section{Introduction}
Seyfert Galaxies, like their more luminous cousins the quasars, contain at their nucleus an accretion disk around a supermassive black hole. As a result of mass accretion into the black hole, their nuclei produce both copious amounts of EUV photons which in turn excite extended narrow-line regions (ENLRs). In addition, energetic bipolar jets of relativistic plasma are produced, which may also shock and energise the ENLR. According to the standard unified model of AGN \citep{Antonucci:90apj,Antonucci:93araa} and its extensions \citep{Dopita:97pasa}, the Seyfert~1 galaxies are seen pole-on relative to the accretion disk, and these display very broad permitted lines originating in rapidly moving gas close to the central engine. In the Seyfert~2 galaxies, the thick accretion disk obscures the central engine, and an ENLR -- often confined within an ``ionisation cone" -- is observed. Within this general paradigm for the formation of the ENLR, many issues remain to be resolved:
\begin{enumerate}
\item {Do Seyfert~2s differ in spatial extent and cone opening angle from the Seyfert~1s \citep{Clarke:98apj,Schmitt:03apjs,Schmitt:03apj}?}
\item {What is the relative energy flux in the EUV continuum compared with the jets, and are the radio-loud objects (which clearly have relativistic jets) more kinematically disturbed \citep{Bicknell:98apj,Evans:99apj, Wilson:99apj}?}
\item {What mechanisms control the non-thermal EUV spectrum, and can these be constrained by observation \citep{Bland-Hawthorn:97apss,Allen:99apj,Done:12mnras,Jin:12a,Jin:12b,Jin:12c}}?
\item{What is the ionisation parameter in the ENLR, how well can this be constrained by the coronal lines \citep{MullerSanchez:11apj} or by other line ratio diagnostics?}
\item{Are all ENLRs dominated by radiation pressure acting on dust \citep{Dopita:02apj,Groves:04apjs, Groves:04apj}?}
\item {What is the chemical abundance distribution in Seyfert nuclei, how well can this be constrained by observations of the surrounding \ion{H}{ii} regions \citep{Evans:87apj}, and how is this correlated with the host galaxy mass? }
\item {To what extent is Seyfert activity triggered by tidal interactions between galaxies, and what is the role of mergers in feeding the AGN, and producing nuclear starbursts?}
\end{enumerate}

To address and understand these issues, especially items 3--6 (above), we have undertaken the {Siding Spring Southern Seyfert Spectroscopic Snapshot Survey} (S7). The S7 survey is an integral field survey in the optical of over 100 southern Seyfert galaxies. The survey uses the Wide Field Spectrograph (WiFeS) mounted on the Nasmyth focus of the ANU 2.3m telescope \citep{Dopita:2010aa}, and will be described in detail in another paper. However, here we present detailed WiFeS observations of the Seyfert~2 galaxy NGC~5427 with the particular objective of providing answers to the questions 3 -- 6 (above) for this galaxy.

NGC~5427 is a giant Sc-type spiral located at a distance of $\sim 40$~Mpc (corresponding to a spatial scale of 177~pc~arcsec$^{-1}$). It contains a Seyfert 2 nucleus \citep{Veron-Cetty:2006aa}. Together with NGC~5426, it forms the interacting pair Arp~271. The interaction between NGC~5427 and NGC~5426 at 20~kpc separation has resulted in a tidal bridge between both galaxies, which is seen in the UV and in H$\alpha$ \citep{Evans:1996aa, Smith:2010aa}. 
The gas kinematics in the interacting pair has been studied using H$\alpha$ kinematical maps obtained with Fabry Perot systems \citep{Fuentes-Carrera:2004aa, Hernandez:2008aa, Font:2011aa}. \cite{Font:2011aa} find evidence for a transfer of hydrogen gas from NGC~5426 to NGC~5427 as well as for a galactic wind in NGC~5427. \cite{Fuentes-Carrera:2004aa} present a detailed study of the orbital configuration of the interaction between NGC~5427 and NGC~5426, providing an estimate of the total dynamical mass of both galaxies within D$_{25}$/2 of $6.72$-$11.2\times 10^{10}\ M_\odot$ for NGC~5426  and $4.5$-$7.5\times 10^{10}\ M_\odot$ for NGC~5427, suggesting that NGC~5426 is the more massive galaxy of the pair within the optical radius.

The host galaxy of NGC~5427 has been classified as SA(s)c pec by \cite{de-Vaucouleurs:1991aa} and is seen nearly face-on. Based on ellipse fitting in the $B$-band, \cite{Marinova:2007aa} report an inclination of $i= 38\deg$ and a position angle of $PA = 11\deg$, although the tidal interactions may make these values rather uncertain. The galaxy stellar mass is estimated to be $M_\star = 4.61\times 10^{10}\ M_\odot$ \citep{Weinzirl:2009aa}, based on the $M_\star$-$(B-V)$ relation from \cite{Bell:2003aa}. Using an SED model for the global UV, optical, and far-infrared (far-IR) emission of NGC~5427, \cite{Misiriotis:2004aa} derive a star-formation rate of $\log \mathrm{SFR} = 0.95\ M_\odot\ \mathrm{yr^{-1}}$, a dust mass of $\log M_\mathrm{d} = 7.83\ M_\odot$ and report a gas mass of $\log M_\mathrm{g} = 10.38\ M_\odot$.

NGC~5427 has been variably classified as barred or un-barred \citep{Fuentes-Carrera:2004aa, Marinova:2007aa, Weinzirl:2009aa, Comeron:2010aa}.
 It has a nuclear ring or pseudo-ring \cite[][and references therein]{Comeron:2010aa} as well as 4-5 nuclear dust spirals in the inner 700~pc \citep{Martini:2003aa}. The circum-nuclear ring of \ion{H}{ii} regions is at a distance of about 1~kpc from the nucleus in H$\alpha$ images \citep{Evans:1996aa, Gonzalez-Delgado:1997aa}. 

NGC~5427 hosts a $\sim 10^7\ M_\odot$ super-massive black hole. By modelling the width of ionised gas emission lines in HST spectra, \cite{Beifiori:2009aa} derive upper limits for the black-hole mass of NGC~5427 of $8.1\times 10^7\ M_\odot$ or $2\times 10^7\ M_\odot$ for a Keplerian disk model assuming inclinations of $i=33\deg$ and $i=81\deg$, respectively. If the accretion disk inclination was the same as the disk inclination of  $i=38\deg$ \citep{Marinova:2007aa}, then the BH mass would be   $\sim 7\times 10^7\ M_\odot$. \cite{Woo:2002aa} estimate a black hole mass of $2.45\times 10^6\ M_\odot$, using the $M_\mathrm{BH}$-$\sigma_\star$ relation from \cite{Tremaine:2002aa}  and a stellar velocity dispersion of $\sigma_\star = 74\ \mathrm{km\ s^{-1}}$. However, this is an indirect method, and therefore subject to greater uncertainty. They also report a bolometric luminosity of $\log (L_\mathrm{bol}) = 44.12\ \mathrm{erg\ s^{-1}}$, estimated by flux integration over the measured (UV to far-IR) spectral energy distribution (SED). If all this luminosity were due to the black hole itself, it would correspond to an Eddington ratio of $\sim 0.4$ with their black hole mass estimate. However, the \cite{Woo:2002aa} estimate is likely to be an underestimate of the true bolometric luminosity when the EUV contribution is taken into account.

In this paper, we derive a model for the EUV SED and luminosity of the accreting supermassive black hole in NGC~5427 based upon a study of the gas excitation and physical conditions in the ENLR surrounding the AGN in NGC~5427. This model is constrained by chemical abundance measurements derived from \ion{H}{ii} region spectra in the host galaxy.  The paper is organised as follows. In Section \ref{obs} we present details of our observations and the data reduction techniques employed, in Section \ref{results} we present emission maps, observed line ratios, and derive the total abundances estimated from the optical spectra of the \ion{H}{ii} regions. In Section \ref{nuc} we describe our photoionisation analysis technique, which we use to constrain the chemical abundances, pressure and ionisation parameter in the NLR, and to estimate the shape of the EUV spectrum.


\section{Observations and data reduction}\label{obs}
\subsection{The radio core of NGC~5427}
\begin{figure}
   \centering
   \includegraphics[width=0.45\textwidth]{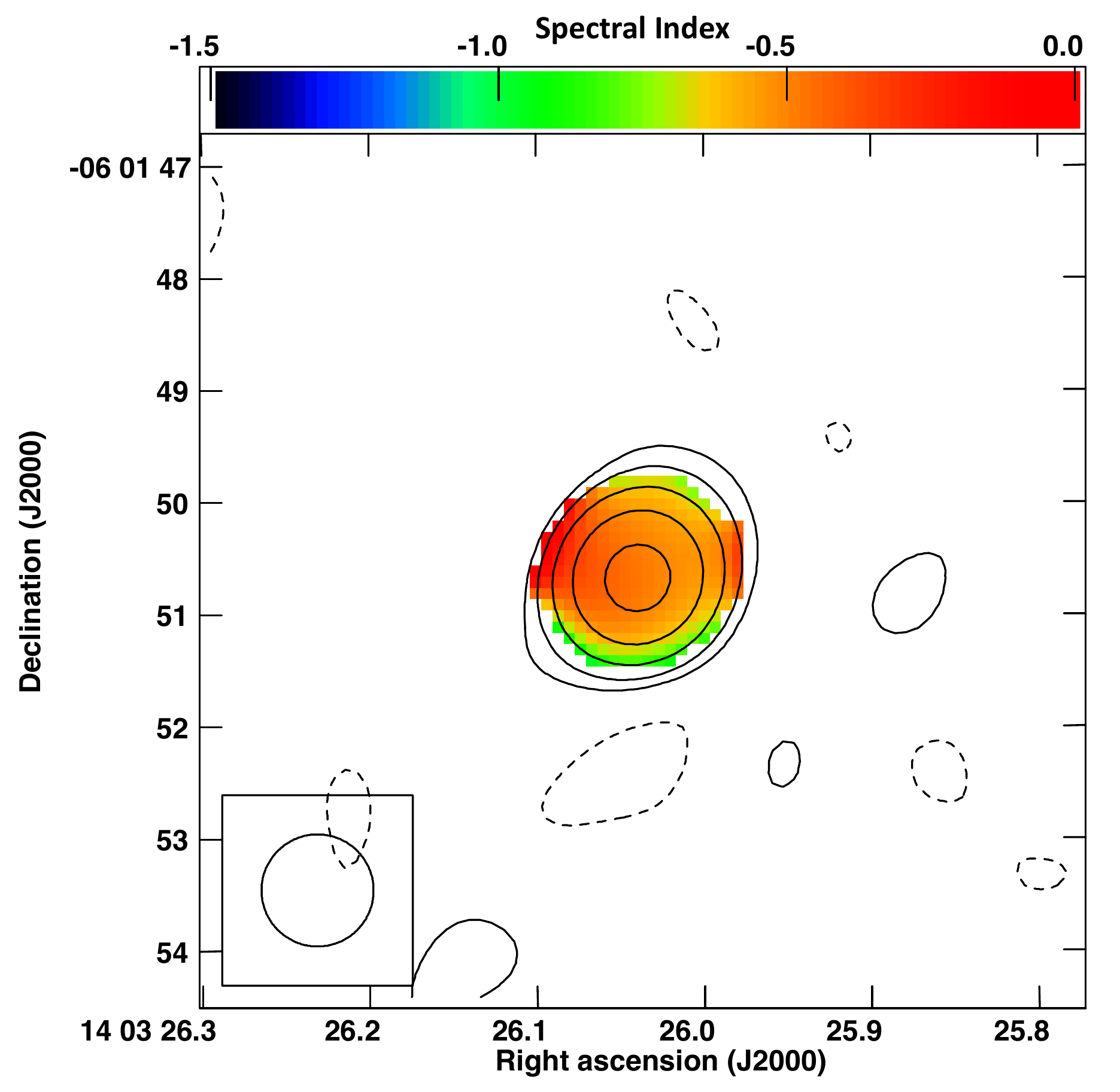}
  \caption{Radio core of NGC~5427 as measured by the VLA A-array at 1.49~GHz (contours are given contours are as a percentage of peak intensity and increase in steps of 2 from the lowest contour of : the peak intensity is 11.8 mJy/beam and the lowest contour of 1.4\% of this. A negative contour of -1.4\% of peak intensity is also shown as a dashed line). The 1.49 - 4.86 GHz spectral index is indicated by the colour scale. The source is essentially unresolved, but there is marginal evidence for an elongation in the NW-SE direction. The spectral index is also steeper in these directions suggesting that the very weak jet within 0.2 arc sec of the nucleus is suffering strong synchrotron losses in its passage through the circum-nuclear ISM. }
  \label{fig:radio_nucleus}
  \end{figure}

We reduced VLA A-array configuration archival data at 1.49 and 4.86 GHz present in the NRAO archive (Project ID: AK212). The data reduction was carried out following standard calibration and reduction procedures in the Astronomical Image Processing System ({\tt AIPS}). Images were made after several iterations of phase and amplitude self-calibration using the {\tt AIPS} tasks {\tt CALIB} and {\tt IMAGR}. The final root mean square (RMS) noise in the images was $\sim8\times10^{-5}$~Jy~beam$^{-1}$ and $\sim6\times10^{-5}$~Jy~beam$^{-1}$ at 1.49 and 4.86~GHz, respectively. At 4.86~GHz, an essentially unresolved radio core is observed with a peak intensity of of 2.0~mJy~beam$^{-1}$. However, at 1.49~GHz, a slightly extended core of 2.0~mJy~beam$^{-1}$ is observed.  The deconvolved size of this as derived by JMFIT is $0.26\arcsec\times0.11\arcsec$; equivalent to 49~pc$\times$21~pc. After constraining the {\tt UVRANGE} to lie between 10$-$185~k$\lambda$ at both the frequencies, and convolving the images with identical beams of size $1\arcsec\times1\arcsec$ (an intermediate resolution between 1.49 and 4.86 GHz) in task {\tt IMAGR}, we created the 1.5$-$4.9 GHz spectral index image using the task {\tt COMB}.  The nuclear source and its spectral index are shown in Figure \ref{fig:radio_nucleus}.

We find that the core has a relatively steep spectrum of $-0.5\pm0.1$, consistent with optically-thin synchrotron emission. Steep spectrum radio cores are fairly prevalent in Seyfert galaxies on both arcsecond- and milliarcsecond-scales  \citep{Sadler:1995mnras, Orienti:2010mnras,Kharb:2010mnras,Panessa:2013mnras}. In NGC~5427 it is likely that the nuclear source is dominated by jet emission, with an elongation in the NW-SE direction. The spectral index is also steeper in these directions suggesting that this weak jet is suffering strong synchrotron losses. The direction of elongation corresponds to the direction along which more extended ``composite'' excitation is seen in the optical (see Figure \ref{fig:nuc_zoom}).

\subsection{The optical observations}
NGC~5427 was observed on 2010 June 12-14 using the Wide Field Spectrograph \citep[WiFeS;] []{Dopita:2007aa,Dopita:2010aa} at the ANU 2.3m telescope at Siding Spring Observatory. WiFeS is an optical integral field spectrograph providing a field of view of 25\arcsec\ $\times$ 38\arcsec\  via 25  38\arcsec\ $\times$ 1\arcsec\ slitlets. The instrument has two arms, one for the blue and one for the red part of the spectrum, so that a wide wavelength range in the optical is covered simultaneously, with adequate spectral overlap between the red and blue spectra. The observations for NGC~5427 were obtained using the $R_S = 3000$ grating in the blue ($B3000$ grating) and the $R_S = 7000$ grating in the red arm ($R7000$ grating).

\begin{figure}
   \centering
   \includegraphics[width=0.5\textwidth]{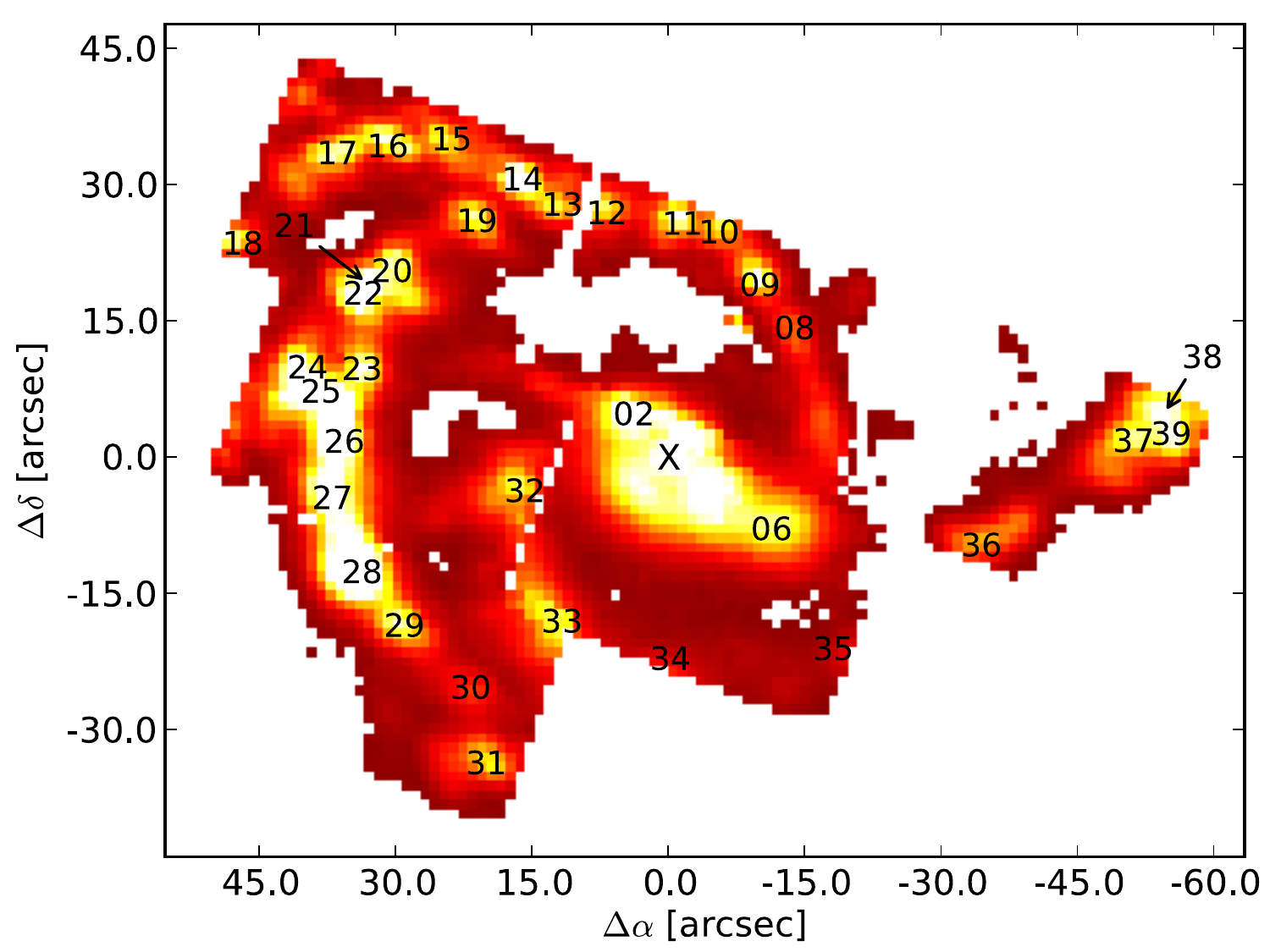}
  \caption{Composite of the six fields observed for NGC~5427 showing the main features in H$\alpha$ at an arbitrary absolute flux scale. The spatial location of the \ion{H}{ii} regions (numbered 02 -- 39) which were used for determining the oxygen abundances, $12+\log(\mathrm{O/H})$, is indicated (see Table~\ref{table_1}).}
  \label{fig:HIIregions}
  \end{figure}

NGC~5427 was observed in six fields covering a large part of the extended disk and the inner spiral arms, see Figure~\ref{fig:HIIregions}.  The observations were performed in nod-and-shuffle mode, which co-adds 6 cycles of 100~s exposures on the object and the same on the nearby sky. These are read out of the CCD at the completion of the sequence. On the chip, the object and the sky data are interleaved. Each field was observed three times in order to be able to remove cosmic rays via median combination of the three individual frames. The exception is the field north to the nuclear data cube (containing \ion{H}{ii} regions 08, 09, 10, 11, 12 in Figure~\ref{fig:HIIregions}), which was observed only twice. Here the two frames were combined after applying cosmic ray rejection based on {\tt L.A.Cosmic} \citep{2001PASP..113.1420V}.

The data were reduced using the newly-released WiFeS Python data reduction pipeline \citep{Childress:2013aa}. The bias subtraction is based on a single bias frame observed close in time to the science data in order to take into account any temporal variations in the bias.To improve the noise inherent to using only one bias frame, we modelled each row in each  of the four quadrants of this bias as the median value of the row. The data are flat-fielded using dome flat fields only, as the weather conditions did not permit to take any twilight flats. The spatial and wavelength calibrations are based on afternoon wire frames and on Ne-Ar arc frames observed close in time to the science object at night, respectively.

The sky subtraction makes use of the interleaved sky spectra on the CCD resulting from the nod-and-shuffle observations. Furthermore, a correction for the differential atmospheric refraction is applied. The red spectra were corrected for telluric absorption based on reference spectra of  early-type main-sequence $B$ stars observed during the nights at a similar airmass as the science object. 

Spectrophotometric standard stars were observed during the nights and used for the spectrophotometric calibration of the blue and red spectra. As the observing conditions were typically non-photometric, the absolute flux calibration is uncertain. For the results of this paper, we only make use of line flux ratios.

The resulting data cubes for the blue and the red grating overlap in wavelength around the cut-off wavelength of the dichroic at 5600~\AA. For the further processing, the blue and red cubes for each field are merged into a single cube, while maintaining the original (different) spectral resolution in the blue and red part of the spectrum. This step is done without applying any further adjustment to their absolute flux scales. To create a smoother transition from the blue to the red, the wavelength region between 5500-5600~\AA\ is kept as an overlap region, where the merged spectrum is the average of the blue and red spectra while keeping in mind that the line information in this region is made uncertain by averaging two spectra at different spectral resolutions. While combining the blue and red cubes, the spectrum is also corrected for the source redshift of $z=0.008733$ \citep{Theureau:1998aa} in order to perform the subsequent analysis in the source rest frame.

\subsection{Continuum subtraction and emission-line fitting}
The continuum in the merged data cubes for each field in NGC~5427 is fitted and then subtracted over a wavelength range from 3700 to 6800~\AA\ 
using a pipeline of {\tt IDL} modules including adjusted versions of the  Voronoi binning \citep{Cappellari:2003aa}, the Penalized Pixel-Fitting method {\tt pPXF} \citep{Cappellari:2004aa}, and the Gas AND Absorption Line Fitting algorithm {\tt GANDALF} \citep{Sarzi:2006aa}. First, the merged data cube for each field is Voronoi-binned in order to obtain a largely uniform continuum signal-to-noise level of $\sim 20$ probed via the median and robust sigma at 4500-4600~\AA. Following the examples distributed with the {\tt GANDALF} software, we fit the binned spectra by first using {\tt pPXF} while masking all emission lines. The resulting stellar kinematics is then provided as input for the simultaneous fit of absorption and emission lines using {\tt GANDALF}. As template spectra, we make use of the single stellar population synthesis results based on the Geneva stellar evolutionary tracks, provided by \citet{Gonzalez-Delgado:2005aa}. These templates are chosen for their sufficiently wide wavelength range and because of their high spectral sampling of 0.3~\AA. We selected a subset of templates covering metallicities of $Z=0.004$, 0.008, 0.02, and 0.04 and ages of $5\times 10^6$, $2.5\times 10^7$, $1\times 10^8$, $2\times 10^8$, $6\times 10^8$, $9\times 10^8$, $1\times 10^9$, $3\times 10^9$, $5\times 10^9$, and $1\times 10^{10}$~yr and convolve each template to the two different WiFeS spectral resolutions in the blue and red before fitting the data. Together with the template spectra, we fit a multiplicative 8th-order polynomial, in order to better match instrumental continuum variations over the wide wavelength range.

While the Voronoi binning improves the reliability of the continuum fitting, a significant amount of spatial information would be lost  when studying the emission lines in the binned version of the continuum-subtracted spectra. For this reason we do not directly use the {\tt GANDALF} results from the emission-line fitting on the binned spectra. We instead recover the full spatial resolution by subtracting the best-fit continuum from each spatially un-binned spectrum in the original WiFeS data cube. This is done after scaling the level of the best-fit continuum in each bin to the continuum level of each original WiFeS spectrum covered by this bin. The continuum levels are again probed via the median flux density in the range 4500-4600~\AA. The emission lines analysis is then performed on the continuum-subtracted un-binned WiFeS spectra via line fitting or direct flux integration using an adjusted version of the {\tt PROFIT} routine \citep{Riffel:2010ab}. Emission line fluxes are extinction-corrected using the extinction curve from \cite{Wild:2011aa}, $f_{\mathrm{corr}} = f_{\mathrm{obs}} [(\mathrm{H}\alpha/\mathrm{H}\beta)_{\mathrm{obs}}/(\mathrm{H}\alpha/\mathrm{H}\beta)_{\mathrm{int}}]^C$, with $C = 3.22\times  [0.6\ (\lambda_{\mathrm{obs}}/5500.)^{-1.3} + 0.4\ (\lambda_{\mathrm{obs}}/5500.)^{-0.7}]$, with intrinsic Balmer decrements $(\mathrm{H}\alpha/\mathrm{H}\beta)_{\mathrm{int}}$ of 2.86 for host galaxy spectra and 3.10 for the AGN spectrum.

\subsection{Image mosaicing}

A common coordinate system for the six WiFeS fields is only used for display purposes and for measuring distances. The line fitting procedure and the extraction of spectra for analysis is performed on the individual data cubes before alignment. For the assignment of the world coordinate system, we extract \ion{H}{ii} region complexes from our H$\alpha$ maps and use the \ion{H}{ii} region catalogue by \cite{Evans:1996aa} as reference. The \ion{H}{ii} regions are extracted using {\tt SExtractor} \citep{Bertin:96aaps} and the coordinate transformation is computed using {\tt IMWCS} \citep{Mink:97}. For the western-most field on NGC~5427, the coordinate system was adjusted by hand, as the automatic computation failed due to a lack of sources. The final mosaic of all data cubes is created using {\tt SWarp} \citep{Bertin:02}.

\section{Basic results}\label{results}

\subsection{Emission-line ratios}

\begin{figure}
   \centering
   \includegraphics[width=0.45\textwidth]{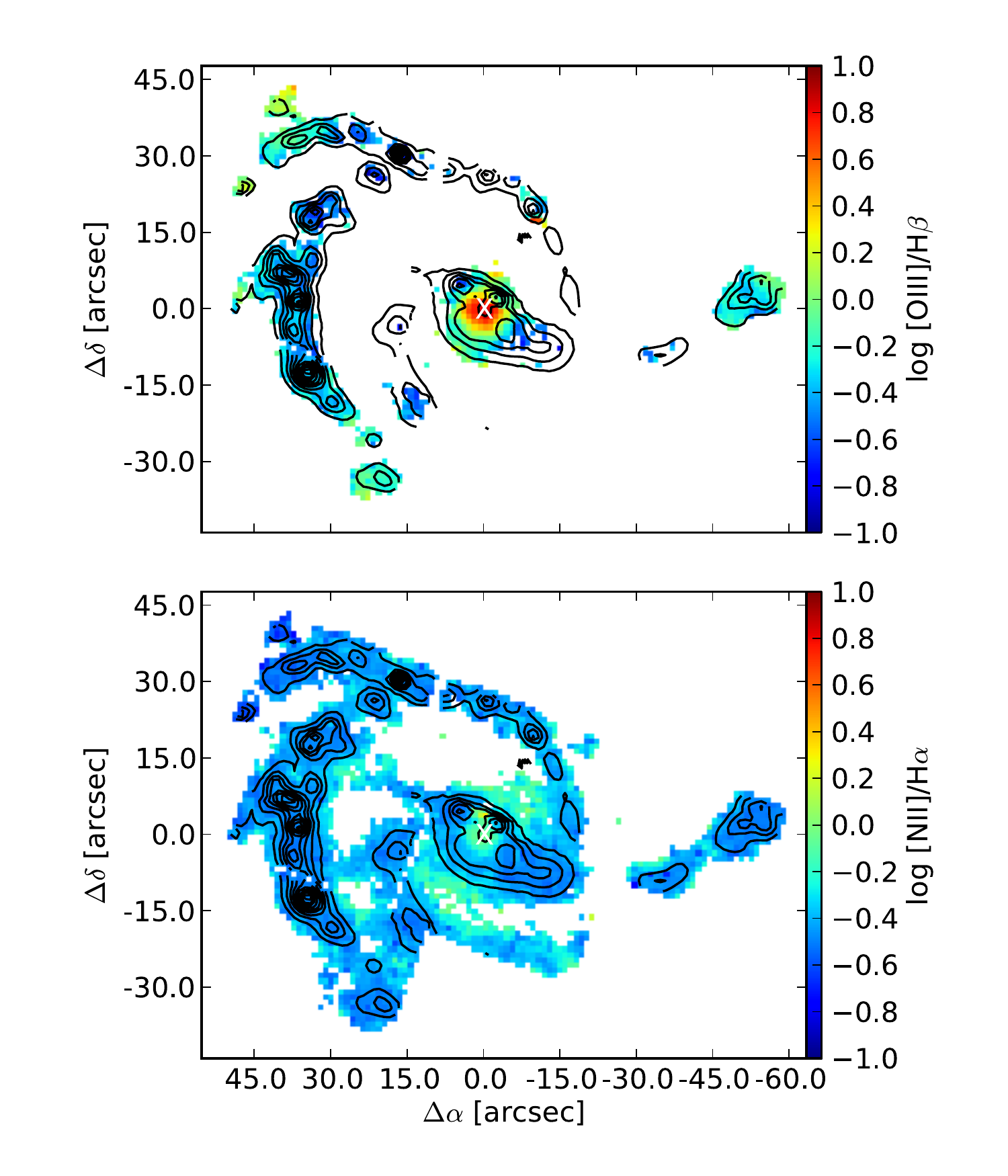}
  \caption{$\log $[\ion{O}{iii}]$\lambda$5007/H$\beta$ and $\log $ [\ion{N}{ii}]$\lambda$6583/H$\alpha$ maps of NGC~5427. Also shown are the H$\alpha$ contours with levels chosen to highlight the main H$\alpha$ features. The data have been clipped using a signal-to-noise ratio of 5 between the line flux and the standard deviation of the nearby continuum multiplied by the FWHM of the line. Note that near the nucleus, the [\ion{N}{ii}]$\lambda$6583/H$\alpha$ ratios are enhanced in the inter-arm regions away from prominent \ion{H}{ii} regions. }
  \label{fig:HaOIIImap}
  \end{figure}

The $\log$[\ion{O}{iii}]$\lambda$5007/H$\beta$ and $\log$[\ion{N}{ii}]$\lambda$6583/H$\alpha$ maps of NGC~5427 maps are shown in Figure~\ref{fig:HaOIIImap} together with H$\alpha$ contours highlighting the main H$\alpha$ features. The data have been clipped using a signal-to-noise ratio of 5 between the line flux and the standard deviation of the nearby continuum multiplied by the FWHM of the line. For computing the signal-to-noise ratio, we used the line fluxes and line velocity dispersions (converted into FWHM) resulting from the profile fitting via PROFIT. 

The maps show the circum-nuclear ring of \ion{H}{ii} regions as well as a large number of bright \ion{H}{ii} regions tracing the spiral arms, in agreement with \cite{Evans:1996aa} and \cite{Gonzalez-Delgado:1997aa}. The H$\alpha$ knots in the spiral arms (in blue tones in this Figure), are largely characterised by small [\ion{O}{iii}]/H$\beta$ and [\ion{N}{ii}]/H$\alpha$ ratios, typical of \ion{H}{ii} regions. In the nuclear region, the [\ion{O}{iii}]/H$\beta$ ratio increases significantly, giving the red tones. This indicates the presence of the central AGN.
 
\begin{figure}
   \centering
   \includegraphics[width=0.45\textwidth]{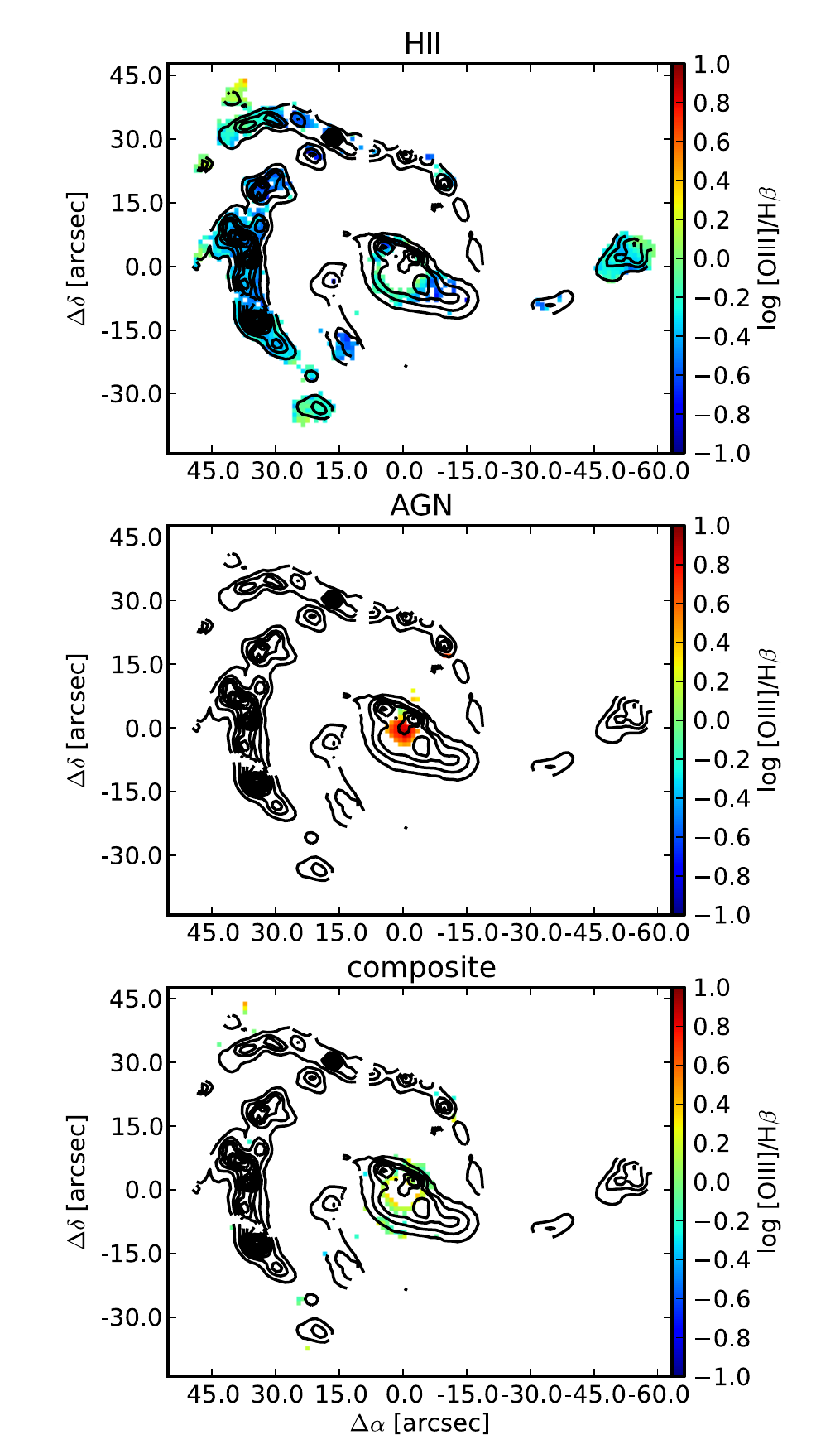}
  \caption{[\ion{O}{iii}]$\lambda$5007/H$\beta$ maps of NGC~5427 showing separately those spaxels that are classified as ``\ion{H}{ii}'', ``composite'', and ``AGN'', respectively. The H$\alpha$ contours are overlaid for comparison. The same clipping as in Figure~\ref{fig:HaOIIImap} is applied.}
  \label{fig:AGNHIIcomp}
  \end{figure}

\begin{figure}
   \centering
   \includegraphics[width=0.5\textwidth]{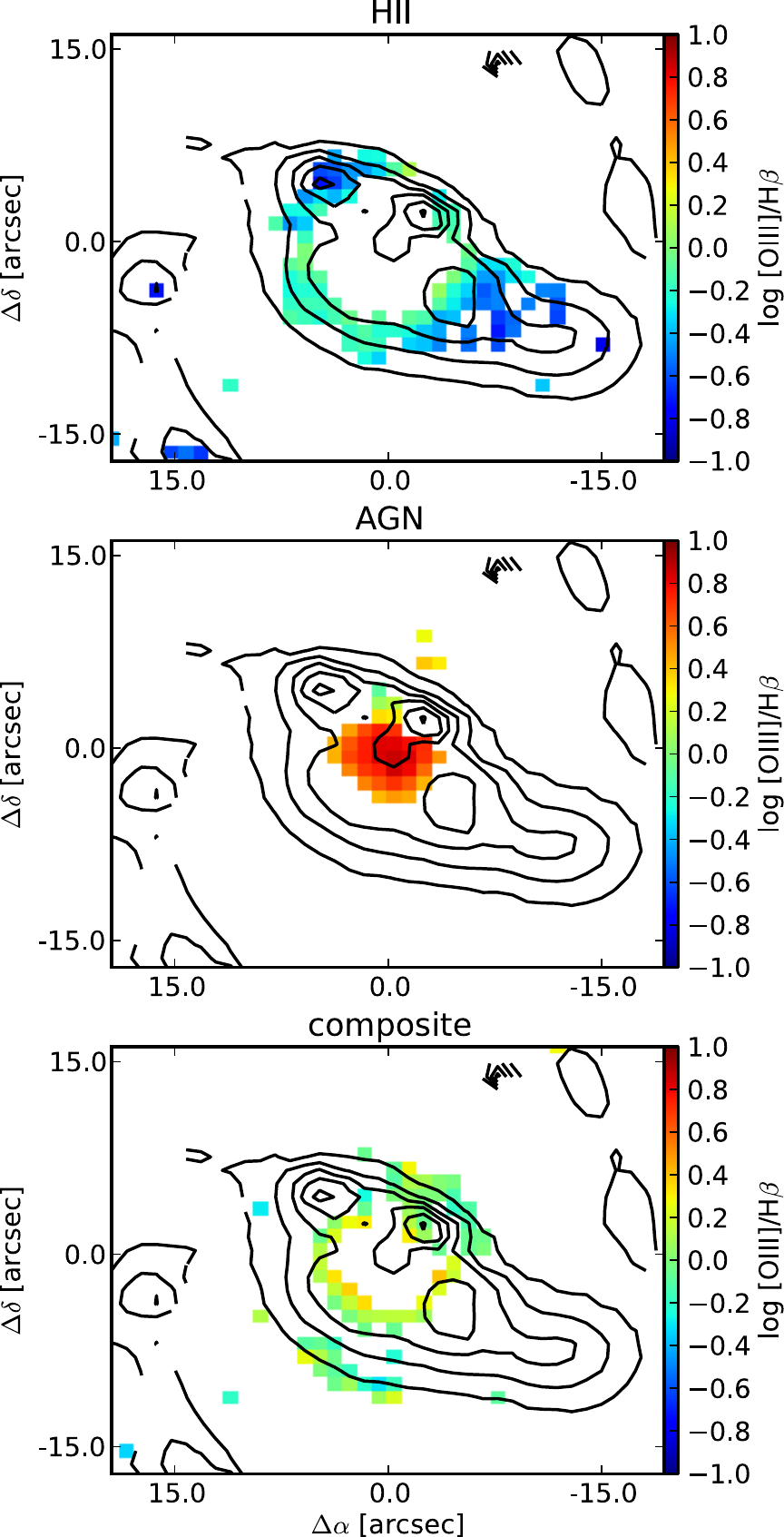}
  \caption{As Figure \ref{fig:AGNHIIcomp}, but zoomed into the nuclear regions. The spatial scale is shown, and the position of the nucleus is marked with ticks. Compare the axis of elongation of the composite points with the 1.49~GHz radio spectral index and intensity map  in Figure \ref{fig:radio_nucleus}.}
  \label{fig:nuc_zoom}
  \end{figure}

There is clear evidence of a weak ENLR. In Figure~\ref{fig:AGNHIIcomp} we show the spatial location of the spaxels separated by their emission-line classification based upon their [\ion{O}{iii}] $\lambda$5007/H$\beta$ and [\ion{N}{ii}]$\lambda$6583/H$\alpha$  ratios. Spaxels are classified as ``AGN'', ``composite'', and ``\ion{H}{ii}"-type, based on the dividing lines defined by \cite{Kewley:2006ab}. For clarity, the nuclear region is zoomed in Figure \ref{fig:nuc_zoom}. As expected, the "AGN"-type line ratios are found in the centre of NGC~5427, while the "\ion{H}{ii}"-type line ratios follow the \ion{H}{ii} regions in the spiral arms. ``Composite"-type line ratios are mostly located in the circum-nuclear region at about the distance of the circum-nuclear ring of \ion{H}{ii} regions. While some of these ``composite" line ratios are an artefact of the AGN point spread function (see the prominent ``ring" like artefact), we identify an extended region of ``composite" line ratios to the south-east and north-west of the nucleus. The gas in these regions is analysed in detail in Section~\ref{sec:composite}, where we argue that this gas traces a contribution from AGN ionisation in the sense of a ENLR or ionisation cone superimposed on distributed emission from normal  \ion{H}{ii} regions. Note that this ENLR is orientated approximately perpendicularly to the flattened inner star forming ring, and in the direction of the radio extension of the nucleus, and the regions of steeper radio spectral index.

 \subsection{Emission-line diagnostics}\label{sec:diagnostics}
  
\begin{figure}
   \centering
   \includegraphics[width=0.45\textwidth]{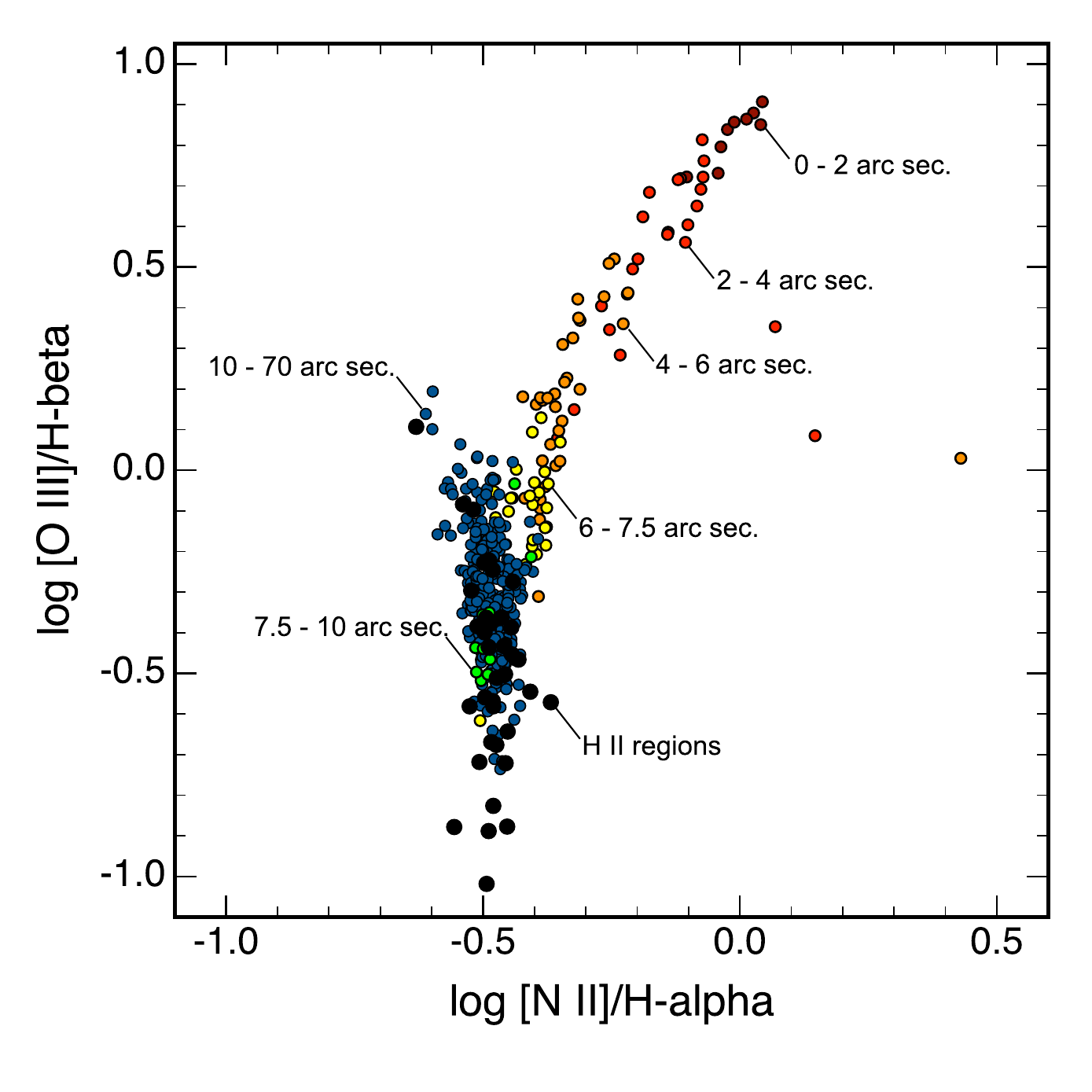}
  \caption{BPT diagram  [\ion{N}{ii}] $\lambda$6584/H$\alpha$ vs. [\ion{O}{iii}] $\lambda$5007/H$\beta$ for NGC5~5427 plotted for individual spaxels (colour) and for integrated \ion{H}{ii} regions (black points). Note how the ``composite'' spaxels are organised according to their distance from the nucleus (indicated on the figure). }
  \label{fig:BPT}
  \end{figure}

 The nature of the ``composite'' spaxels becomes clearer when plotted on the well-known BPT diagram \citep{Baldwin:1981ab} which plots [\ion{N}{ii}] $\lambda$6584/H$\alpha$ vs. [\ion{O}{iii}] $\lambda$5007/H$\beta$; see Figure \ref{fig:BPT}. The data points in this diagram have a Y-shaped morphology   familiar from the corresponding analysis of integrated galaxy spectra \citep[e.g.][]{Kauffmann:2003aa}. ``\ion{H}{ii}"-type spaxels are located on the left-hand side, while a branch of ``composite"-type spaxels connects the "\ion{H}{ii}"-type spaxels with the "AGN"-type. 

There is a clear relationship between the position of the ``composite'' spaxels in Figure \ref{fig:BPT} and the distance of the spaxel from the AGN at the nucleus. For radial distance $> 10\arcsec$ ($>1.77$~kpc), the spaxels join the \ion{H}{ii} region sequence. The  \ion{H}{ii} regions with the highest chemical abundance are located near the bottom of Figure \ref{fig:BPT}, while the lowest abundance points are located both at higher [\ion{O}{iii}] $\lambda$5007/H$\beta$ ratios and further away from the nucleus. This is clear evidence of a global radial abundance gradient in this galaxy.

\subsection{Gas-phase abundances from \ion{H}{ii} regions}\label{sec:HII}

Photoionisation models for strong line analysis of  \ion{H}{ii} regions are now well constrained \citep{LopezSanchez:12mnras,Dopita:2013apjs} so that \ion{H}{ii} region spectra can be used as a tracer of the total (gas plus dust) oxygen abundance. It is important to note that if abundances derived from strong lines are to be compared with abundances derived for the gas phase, the depletion factors listed in \citet{Dopita:2013apjs} must be taken into account. For Oxygen, this lowers the gas-phase abundances by $\sim0.07$~dex.

Here we derive total oxygen abundances for \ion{H}{ii} region complexes in the NGC~5427 data cubes, as shown in Figure~\ref{fig:HIIregions}. We also refer the reader to the H$\alpha$ maps by \citet{Evans:1996aa}.  The \ion{H}{ii} region complexes located closest to the nucleus are excluded from the analysis, as these are contaminated by AGN ionisation and/or the point spread function of the Seyfert nucleus. For these, a simple \ion{H}{ii} region photoionisation modelling would provide incorrect gas-phase abundances. 

For each \ion{H}{ii} region complex, we have extracted a total spectrum from the continuum-subtracted data cubes by summing spectra on the
\ion{H}{ii} region over a $3\times 3$ pixel (i.e. $\sim 3\arcsec \times 3\arcsec $) box. In each of these spectra, the line fluxes are measured by direct integration over the line, whilst subtracting a residual constant baseline determined from the continuum to the left and right of the line. The wavelength window for each line is centred according to the velocity shift determined from a Gaussian fit to H$\beta$ and H$\alpha$ in the blue and red part of the spectrum, respectively. Correspondingly, the size of the wavelength window is set to an appropriate  multiple of the line width of H$\beta$ and H$\alpha$. Both, the Gaussian fit as well as the actual flux integration, are based on adapted versions of the {\tt PROFIT} routine \citep{Riffel:2010ab}. The estimate of the flux error for each line is based on the robust sigma of the nearby continuum. Only the fluxes of lines detected at $> 3\sigma$ are used in the further analysis. The line fluxes are extinction-corrected based on the extinction curve from \cite{Wild:2011aa}, assuming an intrinsic Balmer decrement of 2.86.

The oxygen abundance for each \ion{H}{ii} region is derived from fitting the photoionisation models of \citet{Dopita:2013apjs}  to the following line ratios: [\ion{O}{iii}]/H$\beta$, [\ion{N}{ii}]/H$\alpha$, [\ion{S}{ii}]/H$\alpha$, [\ion{O}{iii}]/[\ion{O}{ii}], [\ion{O}{iii}]/[\ion{N}{ii}], [\ion{O}{iii}]/[\ion{S}{ii}], [\ion{N}{ii}]/[\ion{S}{ii}], and [\ion{N}{ii}]/[\ion{O}{ii}]. We have used the {\sf{pyqz}} code described in that paper with a Maxwell-Boltzmann electron distribution. This delivers a simultaneous solution to the ionisation parameter $\log q$ and the total (gas + dust) Oxygen abundance, 12+$\log$[O/H] using all line ratios that fall upon the diagnostic grids. The error is computed as the RMS error given from all line ratio pairs which provide a physical solution. The results for all \ion{H}{ii} regions shown in Figure~\ref{fig:HIIregions} are listed in Table \ref{table_1}. The  \ion{H}{ii} region Nos. 30 and 35 are not used in the analysis, as the solution for the abundance and ionisation parameter in these cases shows very large scatter, either due to contamination by the ENLR, or due to photometric error (see Table \ref{table_1}. We have also excluded \ion{H}{ii} region No. 2, close and to the north-east of the Seyfert nucleus, because its spectrum is contaminated by broad [\ion{O}{iii}] emission. For this reason, the lower abundance estimated for this \ion{H}{ii} region is certainly an artefact. 
 
\begin{figure}
   \centering
   \includegraphics[width=0.45\textwidth]{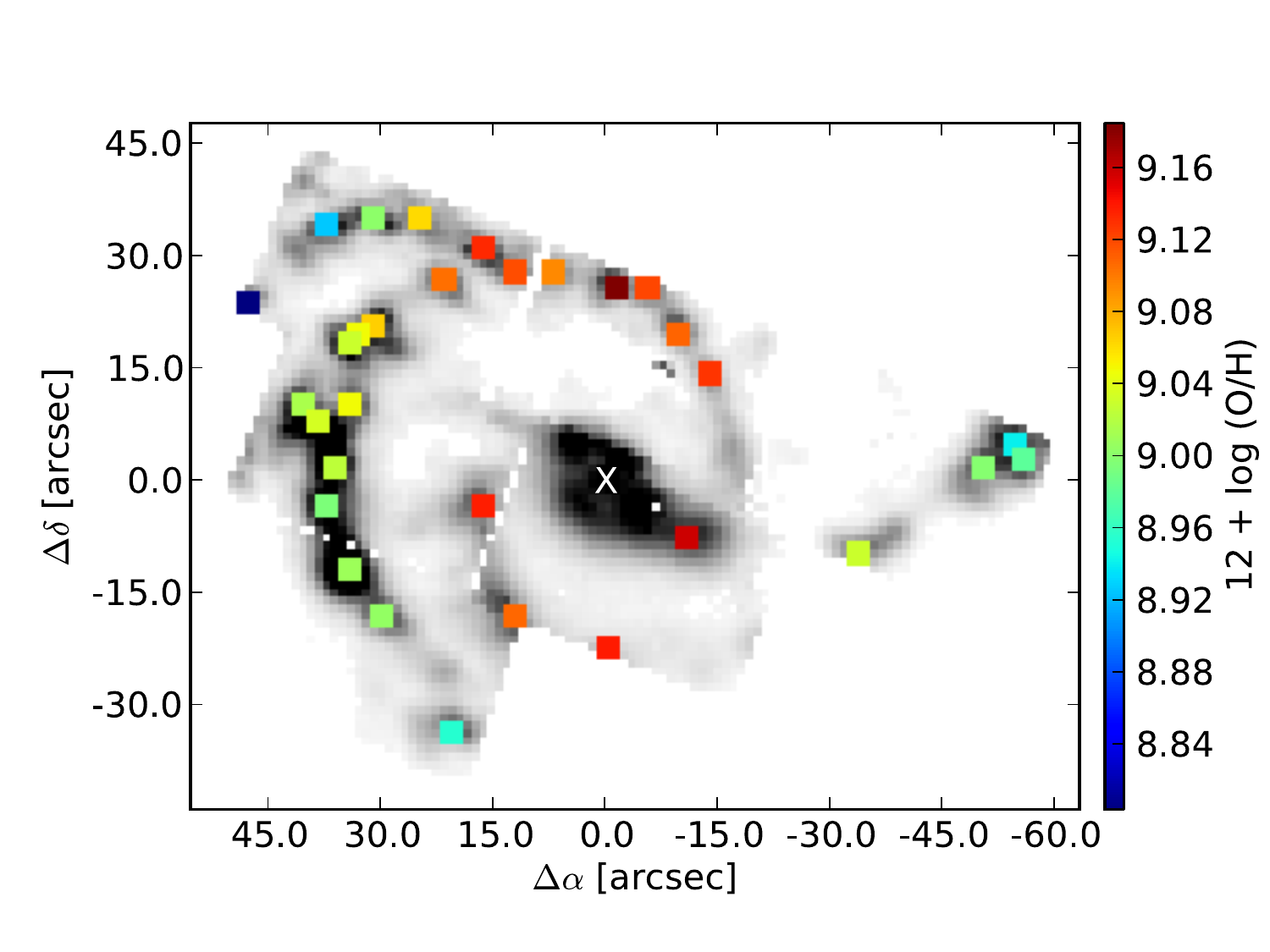}
  \caption{Oxygen abundances in NGC~5427 as derived for \ion{H}{ii} region complexes via photoionisation modelling. 
  The oxygen abundance listed in Table~\ref{table_1} is colour-coded. Region 35 is left out, as the abundance has a large uncertainty. The grey-scale background image shows the H$\alpha$ image in which unreliable measurements have been clipped. The nucleus is marked with a white cross. Note the the presence of an abundance gradient with increasing abundance along the spiral arms toward the central region.}
  \label{fig:HIImetallicity}
  \end{figure}

Figure~\ref{fig:HIImetallicity} shows the resulting oxygen abundances across NGC~5427. It is evident that there is a clear radial abundance gradient in this galaxy. The lowest abundance of $12 + \log (\mathrm{O/H}) = 8.79 (\pm 0.04)$ is found for an \ion{H}{ii} region complex in the outer parts of the spiral arm in the north-east (region 18). The highest abundance of $12 + \log (\mathrm{O/H}) = 9.16 (\pm 0.06)$ is found close to the nucleus to the south-west (region 06). 

\begin{figure}
   \centering
   \includegraphics[width=0.45\textwidth]{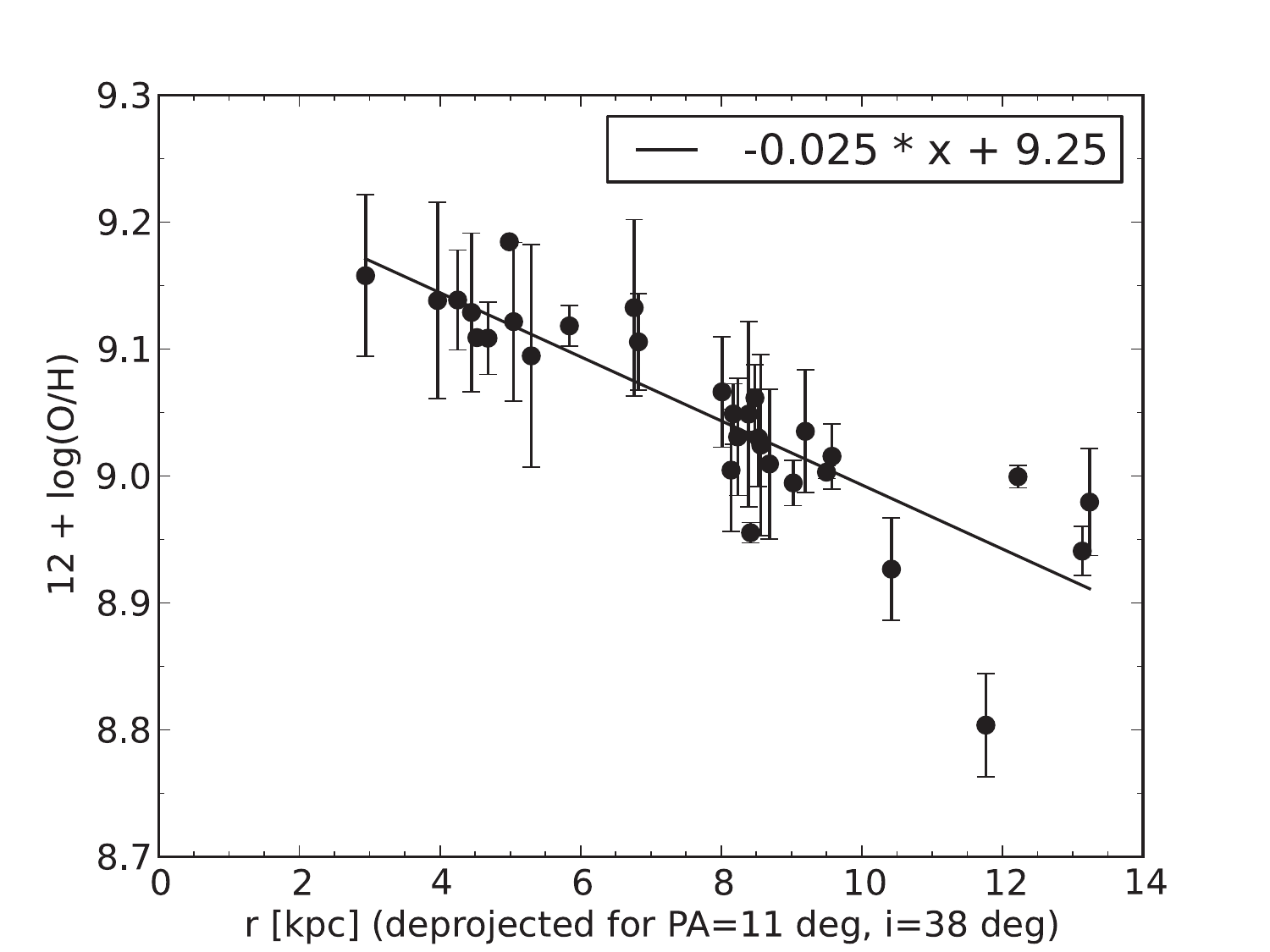}
  \caption{Abundance gradient in NGC~5427. The oxygen abundance (see Figure~\ref{fig:HIImetallicity} and Table~\ref{table_1}) is plotted against deprojected distance
  from the nucleus. Region 35 is left out, as the abundance has a large uncertainty. To compute the deprojected distance, it is assumed that the galaxy disk of NGC~5427 has an inclination of $i=38 \deg$ and a position angle of $PA = 11 \deg$ \citep[cf.][]{Marinova:2007aa}. The line shows a linear regression. The linear regression results in a gradient of -0.025~dex/kpc and a (maximum) central abundance of $12 + \log (\mathrm{O/H}) = 9.25$, typical for galaxies in their early tidal interaction phase.}
  \label{fig:gradient}
  \end{figure}

The de-projected radial gradient of the oxygen abundance in NGC~5427 is fitted with a linear regression in Figure~\ref{fig:gradient}.
This linear regression results in an abundance gradient of $\sim 0.025$~dex/kpc. This measured abundance gradient in NGC~5427 is typical of galaxies in their Stage 1 of the merging process \citep{Rich:12apj}. The abundance gradient extrapolated to the centre implies a nuclear abundance of $12 + \log (\mathrm{O/H}) = 9.25 \pm 0.08$. However, the abundance gradients of galaxies with star-forming rings usually flatten within the ring, so we have conservatively adopted a nuclear abundance  the same as for region No. 06 of $12 + \log (\mathrm{O/H}) = 9.16 \pm 0.08$ (corresponding to a gas-phase abundance $12 + \log (\mathrm{O/H}) = 9.09 \pm 0.08$),or $3.0 Z_{\odot}$ \citep{Grevesse:10apss}. NGC~5427 has a stellar mass of $M_\star = 4.61\times 10^{10}\ M_\odot$ \citep{Weinzirl:2009aa}. With a nuclear metallicity of $12 + \log (\mathrm{O/H}) \sim 9.2$, this galaxy is located at the very metal-rich end of the local mass-metallicity relation of star-forming galaxies \citep[cf.][]{Lara-Lopez:2013aa}. This high abundance is in common with other observational and theoretical studies suggesting high metallicities in Seyfert galaxies \citep{Storchi-Bergmann:1990pasp,Nagao:2002apj,Ballero:2008aap}.

{ We also considered a $ \kappa -$distribution \citep{Nicholls:2012apj,Dopita:2013apjs} (for a more complete discussion of this, see section 4.4). If indeed such a distribution applies to the \ion{H}{ii} regions, the computed abundance gradient with $ \kappa = 20$ is somewhat steeper};  $\sim 0.03$~dex/kpc and the extrapolated central abundance could be as high as  $12 + \log (\mathrm{O/H}) = 9.32$. In general, with $ \kappa = 20$, the computed errors in both $q$ and $Z$ are smaller by a factor of about 1.5-2 than the figures given in Table \ref{table_1}.

\begin{table*}
\centering
\scriptsize
\caption{The logarithms of the observed line ratios for the \ion{H}{ii} regions identified in NGC~5427 (see Figure~\ref{fig:HIIregions}). The mean ionisation parameters, $\log q = \log (cU)$, where $c$ is the speed of light in cm~s$^{-1}$, and the mean total oxygen abundances, 12+log(O/H) are derived from the {\sf pyqz} program (see text). \label{table_1}}
\begin{tabular}{lccccccccccccc}
\hline
& $\log$ Ratios: & & & & & & & & & & & & \\
Region & [\ion{O}{iii}]/H$\beta$ & [\ion{N}{ii}]/H$\alpha$ & [\ion{S}{ii}]/H$\alpha$  & [\ion{O}{iii}]/[\ion{O}{ii}] & [\ion{O}{iii}]/[\ion{N}{ii}] & [\ion{O}{iii}]/[\ion{S}{ii}] & [\ion{N}{ii}]/[\ion{S}{ii}] & [\ion{N}{ii}]/[\ion{O}{ii}] &  $\log$q &  $\Delta \log$q &  12+log(O/H) &  $\Delta \log$(O/H)] \\
\hline
 02$^{1}$ &-0.501&-0.495&-0.664&-0.339&-0.462&-0.293&0.169&0.123&7.39 &0.17& 9.11 &0.06\\
 06 &-0.769&-0.483&-0.736&-0.931&-0.743&-0.49&0.253&-0.188&7.37 &0.29& 9.16 &0.06\\
 08 &-0.803&-0.494&-0.647&-1.017&-0.765&-0.613&0.153&-0.251&7.15 &0.17&9.13 &0.06\\
 09 &-0.489&-0.466&-0.619&-0.651&-0.48&-0.327&0.153&-0.172&7.35 9&0.01& 9.11 &0.01\\
 10 &-0.631&-0.46&-0.656&-0.911&-0.628&-0.431&0.197&-0.282&7.27 &0.22& 9.12 &0.06\\
 11 &-0.906&-0.487&-0.654&---&-0.876&-0.709&0.167&---& 7.27 &0.00& 9.18 &0.01\\
 12 &-0.76&-0.474&-0.602&-1.146&-0.742&-0.615&0.128&-0.404& 7.07 &0.23& 9.09 &0.09\\
 13 &-0.828&-0.517&-0.597&-0.912&-0.767&-0.688&0.079&-0.145&7.09 &0.03& 9.12 &0.02\\
 14 &-0.689&-0.445&-0.65&-0.968&-0.7&-0.496&0.204&-0.268& 7.26 &0.24& 9.13 &0.07\\
 15 &-0.45&-0.455&-0.56&-0.795&-0.451&-0.346&0.105&-0.343&7.21 &0.09& 9.06 &0.03\\
 16 &-0.245&-0.49&-0.539&-0.648&-0.212&-0.163&0.049&-0.436&7.21 &0.01& 9.00 &0.01\\
 17 &-0.113&-0.549&-0.527&-0.503&-0.02&-0.043&-0.023&-0.483&7.23 &0.08& 8.92 &0.04\\
 18 &0.101&-0.615&-0.532&-0.596&0.26&0.177&-0.084&-0.856& 7.16 &0.07& 8.80 &0.04\\
 19 &-0.68&-0.461&-0.583&-0.925&-0.675&-0.554&0.121&-0.249& 7.15 &0.12& 9.11 &0.04\\
 20 &-0.542&-0.495&-0.602&-0.874&-0.504&-0.397&0.107&-0.37&7.17 &0.13& 9.07 &0.04\\
 21 &-0.624&-0.49&-0.569&-1.052&-0.59&-0.512&0.078&-0.462& 7.06 &0.17& 9.05 &0.07\\
 22 &-0.529&-0.493&-0.534&-0.928&-0.492&-0.451&0.041&-0.436&7.06 &0.09&9 9.03 &0.04\\
 23 &-0.471&-0.453&-0.525&-0.836&-0.475&-0.403&0.072&-0.361&7.15 &0.08& 9.05 8&0.02\\
 24 &-0.38&-0.463&-0.509&-0.818&-0.373&-0.327&0.046&-0.445& 7.13 &0.07& 9.02 &0.03\\
 25 &-0.47&-0.457&-0.531&-0.917&-0.469&-0.395&0.074&-0.448&7.12 &0.13& 9.04 &0.05\\
 26 &-0.413&-0.489&-0.591&-0.893&-0.381&-0.279&0.102&-0.512&7.17 &0.18& 9.02 &0.07\\
 27 &-0.37&-0.508&-0.518&-0.785&-0.318&-0.309&0.01&-0.466&7.09 &0.03& 8.99 &0.02\\
 28 &-0.39&-0.49&-0.56&-0.871&-0.356&-0.287&0.07&-0.515&7.13 &0.14& 9.01 &0.06\\
 29 &-0.41&-0.498&-0.543&-0.87&-0.368&-0.324&0.045&-0.502& 7.10 &0.10& 9.00 &0.05\\
 30$^{1}$ &-0.276&-0.439&-0.461&-1.06&-0.293&-0.272&0.021&-0.767&7.02 &0.23& 8.93 &0.13\\
 31 &-0.234&-0.508&-0.494&-0.708&-0.183&-0.196&-0.014&-0.525&7.12 &0.01& 8.96 &0.01\\
 32 &-0.852&-0.494&-0.679&-1.079&-0.814&-0.63&0.184&-0.265&7.18 &0.24& 9.14 &0.08\\
 33 &-0.663&-0.531&-0.664&-0.815&-0.588&-0.456&0.133&-0.226&7.22 &0.07& 9.11 &0.03\\
 34 &-0.602&-0.364&-0.561&-0.864&-0.694&-0.497&0.197&-0.17&7.30 &0.17& 9.14 &0.04\\
 35$^{1}$ &-0.468&-0.469&-0.612&-1.32&-0.456&-0.313&0.143&-0.864&7.08 &0.45& 8.97 &0.22\\
 36 &-0.405&-0.45&-0.534&-0.866&-0.411&-0.327&0.084&-0.455& 7.15 &0.14& 9.03 &0.05\\
 37 &-0.323&-0.524&-0.548&-0.695&-0.255&-0.231&0.024&-0.441& 7.15 &0.01&0 9.00 &0.01\\
 38 &-0.112&-0.526&-0.535&-0.628&-0.042&-0.033&0.009&-0.585& 7.20 &0.03& 8.94 &0.02\\
 39 &-0.255&-0.507&-0.552&-0.749&-0.205&-0.16&0.045&-0.544& 7.16 &0.08& 8.79 &0.04\\
\hline
\end{tabular}
{ \small $^1$ Region of low signal to noise or nuclear contaminated region not used in the abundance gradient analysis.    }
\end{table*}

\section{Analysis of the NLR}\label{nuc}
\subsection{Nuclear spectrum}
\begin{figure}
   \centering
   \includegraphics[width=0.5\textwidth]{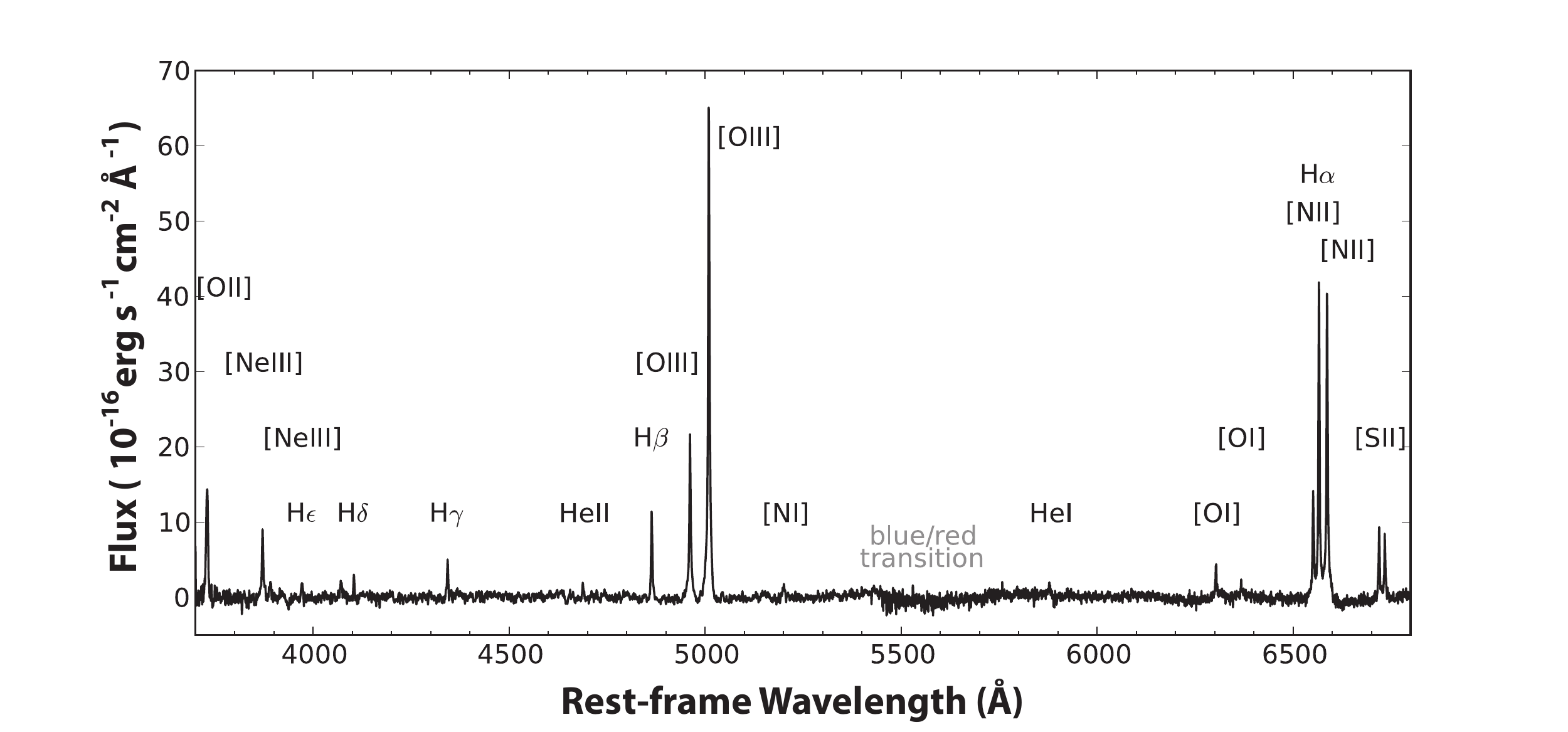}
  \caption{Nuclear spectrum of NGC~5427 at rest-frame wavelengths after correction for Galactic extinction and subtraction of the stellar continuum. The spectrum is summed over a $3\times 3$ pixel (i.e. $\sim 3\arcsec\times3\arcsec$) aperture centred on the nucleus. The principal lines are indicated together with wavelength of  the transition from the blue to the red WiFeS spectrum with their different spectral resolutions (3000 and 7000, respectively). This is a typical Seyfert 2 spectrum. Note that, as mentioned in Sect.~\ref{obs}, the absolute flux scale is uncertain. Based on the available photometric standard stars during the observing night, we estimate the flux uncertainty amounts to a factor of $\sim 1.6$.}
  \label{fig:nuc_spectrum}
  \end{figure}

We now use the gas-phase abundance of the nuclear region obtained from the \ion{H}{ii} region measurements in Section~\ref{sec:HII} as an input parameter for modelling the AGN emission-line ratios in NGC~5427. For the input AGN spectrum, we compute the total spectrum summed over a $3\times 3$ pixel (i.e. $\sim 3\arcsec\times3\arcsec$) region centred on the nucleus of NGC~5427 as defined by the peak in [\ion{O}{iii}]$\lambda$5007 emission. The line fluxes in this spectrum are derived by fitting the lines with a Lorentzian profile. The fluxes are corrected for extinction using the extinction curve from \cite{Wild:2011aa} and assuming an intrinsic Balmer decrement of 3.1, to allow for a collisional excitation component of H$\alpha$. The extracted nuclear spectrum is shown in Figure \ref{fig:nuc_spectrum}.

\subsection{Analysis code}
Here, we have used an upgraded version of the {\tt Mappings IV} code (version 4.1) described by \citet{Dopita:2013apjs} and references therein. This version incorporates the following improvements. 

First, it incorporates a new global predictor -- corrector integration scheme. This, along with improved equilibrium calculations allows us to achieve a more precise balance of heating and cooling (less than 1 part in $10^5$) and in a shorter computation time. 

Second, the photoelectric cross sections for C, N and O have been refit to the \citet{Verner:1996apj} threshold modified Hartree-Slater cross sections, using threshold data of the NORAD atomic database  \citep{Nahar2013}.  The He I cross section has been refit to the experimental data of \citet{Samson1994}.  The  \ion{H}{i} and \ion{He}{ii} photoelectric cross-sections have been slightly revised to exactly fit the recomputed hydrogenic ground cross sections  given by the NORAD atomic database \citep{Nahar2013}. 

Third, hydrogenic free-bound recombination calculations have been revised.  New coefficients were calculated using the {\tt PIXZ1} subroutine of \citet{Storey1991}, and the methodology of \citet{Ercolano:2006mnras}.  The free-bound coefficients for n=2 to n=30 edges, summing over n = 300 levels to ensure convergence.  The recapture to n = 1 levels is computed in MAPPINGS using the Milne relation directly on the ground  level photoelectric cross sections from the photoionisation calculations to ensure detailed balance and energy conservation.

Lastly, the suite of charge exchange reactions has been to updated to include 51 hydrogen and 48 helium charge exchange reactions for both recombination and ionisation \citep{Kingdon:1996apjs}.  With the extended charge exchange network, particular care was taken to make the hydrogen ionisation equilibrium calculations more robust in the case of very low photon ionisation rates as charge exchange rates become dominant instead.

\subsection{Photoionisation modelling}
To model the nuclear spectrum, we have built at set of dusty radiation-pressure dominated photoionisation models, similar to those described by \citet{Dopita:02apj}, \citet{Groves:04apja}, and \citet{Groves:04apjs}. In these, the radiation pressure (acting mostly on the dust) compresses the gas near the ionisation front. Eventually, when the radiation pressure is high enough, a self-similar structure is produced in the low ionisation gas which generates an optical spectrum that is virtually invariant against change in the ionisation parameter, $U$. In the high-ionisation zone the gas pressure dominates, and a coronal region is produced when $\log U \gtrsim -1.0$. Recently, similar models have been generated using the {\tt Cloudy} code of \citet{Ferland:98pasp} by \citet{Stern:13mnras}. These provide further insight into the basic physics of this class of models. 

The emission line spectrum of the narrow line region is determined by three factors, the chemical abundance (and dust depletion) set, the ionisation parameter, $U$, and the shape of the extreme ultraviolet (EUV) to X-ray SED; roughly extending from the Lyman Limit up to about 10~keV (harder photons are unlikely to be absorbed in the narrow line region). The derivation of the nuclear metallicity from the \ion{H}{ii} regions provides an extremely valuable constraint on the modelling, since it effectively eliminates an extra dimension of modelling space.

Although the form of the EUV spectrum cannot be directly observed from either the ground (or at these wavelengths, from space), the optical line emission spectrum provides strong constraints on the EUV SED once the abundance set has been constrained from the measured \ion{H}{ii} region abundances. For a given SED, the line spectrum is then only dependent upon the ionisation parameter, provided that the pressure in the ionised plasma is not sufficient to collisionally de-excite the important forbidden lines. The method we employ here is reminiscent of the energy balance or Stoy technique used to estimate the effective temperature of stars in planetary nebulae \citep{Stoy1933,Kaler1976,Preite-Martinez1983}, since as the radiation field becomes harder, the heating per photoionisation increases, and the sum of the fluxes of the forbidden lines becomes greater relative to the recombination lines. However, individual line ratios are sensitive in different ways to the form of the EUV spectrum. Here we briefly identify and review the key sensitivities.
\begin{itemize}
\item {He I $\lambda 5876$/H$\beta$ ratio: This directly measures the ratio of the number of photons above 24~eV to the total number of ionising photons. It is therefore sensitive to the form of the ``big blue bump'', and its ratio to the intermediate component. The hard X-ray component is not so critical, as the dust and gas opacity falls rapidly with energy.  With radiation pressure dominated models (at high $U$), the absorption by dust of the ionising photons becomes important \citep{Dopita:2006apj} and the He I $\lambda 5876$/H$\beta$ ratio becomes dependent upon the ionisation parameter as well.}
\item{He II $\lambda 4686$/He I $\lambda 5876$ ratio: Fundamentally, this should be sensitive to the ratio of the photons above 54~eV and above 24~eV. However, this ratio is also sensitive to $U$ owing to the competition of dust in absorbing the softer H- and He-ionising photons.}
\item{ [\ion{O}{iii}] $\lambda 5007/$H$\beta$ and [\ion{Ne}{iii}] $\lambda 3868/$H$\beta$ ratios: These are primarily sensitive to $U$. However, they can become very large in the presence of an intermediate Comptonised component. The [\ion{O}{iii}] $\lambda 5007/$H$\beta$ ratio in particular provides a stringent constraint on the intensity of this component.}
\item{ [\ion{O}{i}] $\lambda 6300/$H$\alpha$,  [\ion{N}{i}] $\lambda 5200/$H$\beta$ and  [\ion{S}{ii}] $\lambda\lambda 6717,31/$H$\alpha$ ratios: These are all strongly dependent on the degree of the X-ray heating, as they arise from the partially ionised tail near the ionisation front heated by Auger electrons. These ratios strongly constrain the form of the X-ray spectrum, but they also become sensitive to $U$ once the radiation pressure exceeds the gas pressure (roughly, at $\log U > -2.5$).}
\item{ [\ion{S}{ii}] $\lambda$ 6717/[\ion{S}{ii}] $\lambda$6731 ratio: This determines the pressure in the recombination zone (which is the sum of the initial gas pressure, and the pressure in the radiation field which has been absorbed up to that point). In the case of NGC~5427 the  [\ion{S}{ii}]  ratio indicates a density of $\sim 350$~cm$^{-3}$. In our models this density was kept constant by adjusting the initial gas pressure to the appropriate value for each value of the ionisation parameter, $U$. }
\end{itemize}
Together these provide a set of independent constraints on the form of the ionising spectrum from the Lyman limit up to an energy of several keV.

\subsection{Fitting the nuclear emission line spectrum}
The majority of modelling of the NLR spectrum hitherto has been done with simple power-law photon distributions; see \cite{Stern:13mnras} and references therein. This approach lacks physical justification, and does not properly account for the EUV shape of the ``big blue bump'' which is the signature of the accretion disk. This can be done because the standard \citet{SS:73aap} $\alpha$-disk is only a rather weak and soft producer of EUV photons.

A more physically-based model for the  SED of Seyfert galaxies has recently been provided by Done, Jin and their collaborators \citep{Done:12mnras,Jin:12a,Jin:12b,Jin:12c}. In these models, the EUV spectrum consists of three components, a big blue bump extending out to around 100~eV, an intermediate Comptonised component dominating between 300~eV and 2~keV (roughly), and a hard X-ray component extending from around 100~eV to above 100~keV. In these models, the absorption opacity of the inner disk is explicitly solved resulting in a somewhat harder accretion disk spectrum. The intermediate  Comptonised component was invoked to explain the soft X-ray excess which is frequently observed. However, this may or may not be present depending on the presence and distribution of gas close to the black hole above and below the disk.

 Our modelling quickly revealed that such an intermediate Comptonised component could not be present at an appreciable level in the source spectrum. Such a component would cause both the He II $\lambda 4686$/He I $\lambda 5876$ ratio and the [\ion{O}{iii}] $\lambda 5007/$H$\beta$ ratio to far exceed their observed values. Also the  [\ion{O}{i}],  [\ion{N}{i}] and [\ion{S}{ii}] lines would be predicted to be much stronger than observed. This component in any case was predicted to arise by re-processing in the gas surrounding the AGN, for which we have little evidence here. Indeed, NGC~5427 has been proposed \citep{Ghosh:07apj} to be a ``true'' Seyfert 2, in that the BLR is not ``hidden'' in this source as per Unification, but is more likely absent. This lack of obscuration was deduced from optical and UV HST data and Chandra observations of this source. They find no hard X-ray emission and very minimally extended soft X-ray emission. We therefore eliminated the intermediate-energy Comptonised component from the \citet{Done:12mnras} SED and simply investigated the effect of scaling the hard X-ray power-law component relative to the big blue bump (accretion disk) component. The ionisation parameter $U$ was treated as a free variable, and varied in steps of 0.25~dex in the range $-3.0 \leqslant \log U \leqslant 0.0$.
 
The strength of the [\ion{O}{iii}] , [\ion{O}{i}], and [\ion{S}{ii}] lines relative the the H I recombination lines, and the He II $\lambda 4686$/He I $\lambda 5876$ ratio constrains the relative strength of the hard X-ray component rather strongly. We ran four sets of models with hard X-ray scale factors of 0.2, 0.25, 0.3, and 0.35. Below a scaling factor of 0.2, it is not possible to obtain a fit to these ratios, and above 0.3 the quality of the best-fit to the spectrum declines rapidly. We also investigated the effect of varying the input mass and Eddington fraction of the black hole. The best fit is obtained for $M_{\rm BH} = 5\times 10^7 M_{\odot}$ and an Eddinton fraction of 0.1, but an acceptable fit can also be obtained with $M_{\rm BH} =10^8 M_{\odot}$ and an Eddington fraction of 0.1. Black hole masses as small as $M_{\rm BH} =10^6M_{\odot}$ are definitely excluded, as the big blue bump component  becomes too hard.
 
\begin{figure*}
 \centering
   \includegraphics[width=0.9\textwidth]{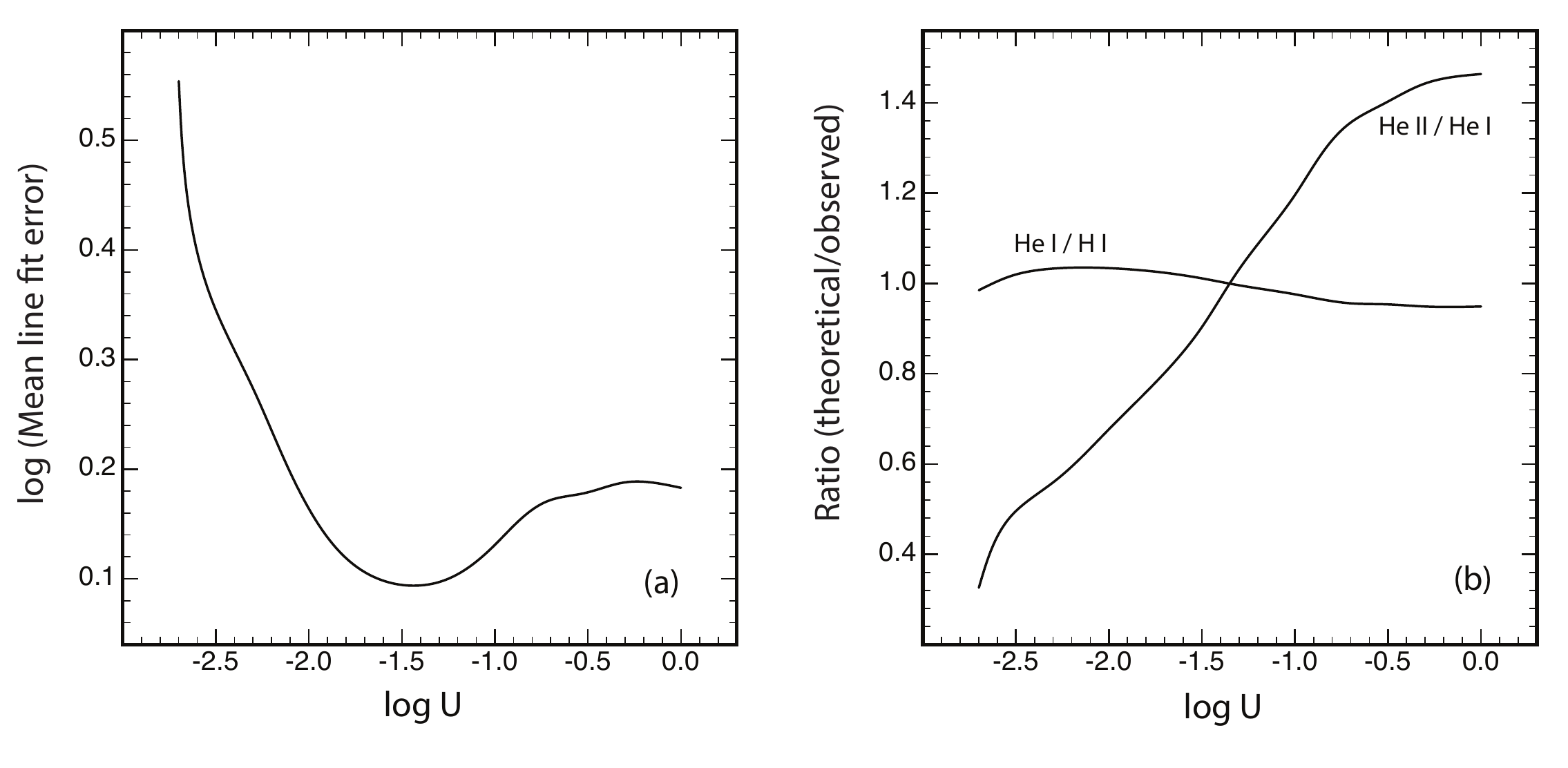}
  \caption{Determination of the best fit to the nuclear spectrum of NGC~5427 (model A) for $M_{\rm BH} = 5\times 10^7 M_{\odot}$, an Eddington fraction of 0.1 and a hard X-ray scale factor of 0.2. The left panel (a) shows the RMS error (in logarithmic space for 13 emission  lines, while the right panel (b) compares the \ion{He}{ii} $\lambda 4686$/\ion{He}{i} $\lambda 5876$ ratio and the \ion{He}{i} $\lambda 5876$/H$\beta$ ratio with the observed ratios. It is evident that the best fit in this case is obtained at $\log U \sim -1.3$. The line fluxes of this  fit are given in detail in Table \ref{table_2}.}
  \label{fig:fitting}
  \end{figure*}

Our fitting procedure was to minimise (in logarithmic space) the RMS difference between the model and the observational line flux ratio (with respect to H$\beta$) for 13 lines simultaneously. Working in logarithmic space ensures that the fainter lines are weighted in the same way as the brighter lines. The lines used in the fitting were  [\ion{O}{ii}] $\lambda\lambda$ 3727,9,  [\ion{Ne}{iii}] $\lambda 3868/$, H$\gamma$, \ion{He}{ii} $\lambda 4686$, H$\beta$, [\ion{O}{iii}] $\lambda 5007$, \ion{He}{i} $\lambda 5876$,  [\ion{O}{i}] $\lambda 6300$ and $\lambda 6363$, H$\alpha$,  [\ion{N}{ii}] $\lambda 6583$ and $\lambda 6548$,  [\ion{S}{ii}] $\lambda$ 6717 and [\ion{S}{ii}] $\lambda$6731. In addition, we compared the  \ion{He}{ii} $\lambda 4686$/\ion{He}{i} $\lambda 5876$ ratio and the \ion{He}{i} $\lambda 5876$/H$\beta$ ratio with the observed ratios. When these two ratios are matched to the observations, we can be sure that we have the correct ratio of H-ionising, \ion{He}{i}-ionising and \ion{He}{ii}-ionising photons. The models have initial pressures chosen such that they all generate the correct electron density in the [\ion{S}{ii}] - emitting zone (330 cm$^{-3}$). The resultant fit for the best-fit case of $M_{\rm BH} = 5\times 10^7 M_{\odot}$, an Eddington fraction of 0.1 and a hard X-ray scale factor of 0.2 are shown in Figure \ref{fig:fitting}.

To obtain such a good fit to the measured emission-line ratios, the He abundance needed to be enhanced above that found from the \ion{H}{ii} regions by +0.16~dex, and the N abundance by +0.3~dex. Similar evidence of N enhancement in the nuclear regions of Seyfert galaxies have already been inferred by other authors \citep[e.g.][]{Storchi-Bergmann:1991aa, Dopita:1995aa, Scharwachter:2011aa}. 
 
Enhanced N and He is an indication of CN-processed gas, which (we may speculate) may be produced by massive, rapidly rotating stars, enriching the ISM with the N-rich partially-burnt CNO products through their stellar winds. The data therefore suggest that such stars may be embedded within the more spatially extended AGN accretion disk, a 10-100 pc structure referred to as a Q-disk or ``marginally stable stationary disk'' by  \citet{Collin:99aa,Collin:08aa}. This is distinguished from the inner disk by the nature of the viscosity within it.  Stars being formed and embedded in this more extended accretion region could provide a key source of viscosity through their stellar winds and their supernova explosions. However, being embedded within a dense, dusty extended ($\sim 100$pc) disk structure, such stars would be surrounded only by dust-obscured ultra-compact \ion{H}{ii} regions, and would be very hard to detect at optical wavelengths. However, they might contribute to the IR signature of the disk. If confirmed in other objects, this tentative evidence of massive star formation in the extended ($\sim 100$pc) outer disk structure could provide the solution to the ``angular momentum problem'' whereby the orbiting gas around the central black hole can flow in from $\sim 100$~pc down to parsec scales.
 
  \begin{figure*}
 \centering
   \includegraphics[width=0.9\textwidth]{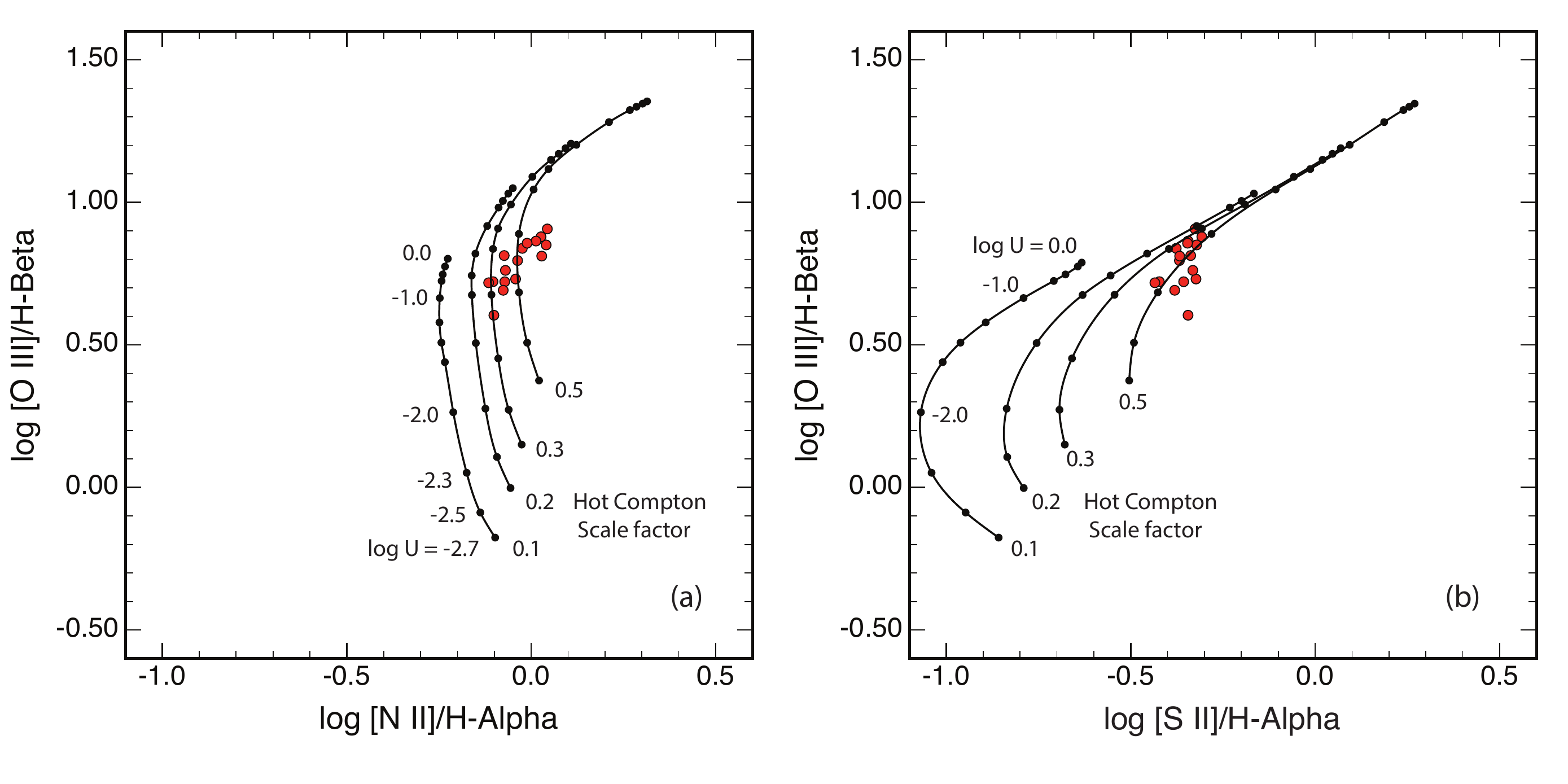}
  \caption{BPT diagrams for the individual spaxels within 2.5 arc sec of the nucleus (red points). We show four sets of models with ionisation parameter in the range $-2.7 \leqslant \log U \leqslant 0.0$, and with different scaling of the hard X-ray component relative to the accretion disk component (0.1 up to 0.5). Note the degeneracy of the solutions on this diagram; good fits to the nuclear line ratios on a BPT diagram can be obtained with different pairs of $U$ and X-ray scaling factor. However, the excitation of helium breaks this degeneracy; see Figure \ref{fig:fitting}.
  }
  \label{fig:fit}
  \end{figure*}

With this change in N abundance, a good fit can also be obtained on the classical BPT diagrams; see Figure \ref{fig:fit}. However, in these diagrams there is a degeneracy between the hard X-ray scaling factor and the ionisation parameter $U$. Harder spectra are fit with lower $U$. However, as the X-ray scaling factor is decreased, eventually a solution becomes no longer possible because the [\ion{O}{iii}]/H$\beta$ ratio saturates at a maximum value, which is less than the observed value. The strongest observational constraint on the hard X-ray scaling factor comes from the \ion{He}{ii} $\lambda 4686$/\ion{He}{i} $\lambda 5876$ ratio and the \ion{He}{i} $\lambda 5876$/H$\beta$ ratio; \emph{c.f.} Figure \ref{fig:fitting}.
 
\begin{figure}
 \centering
   \includegraphics[width=0.45\textwidth]{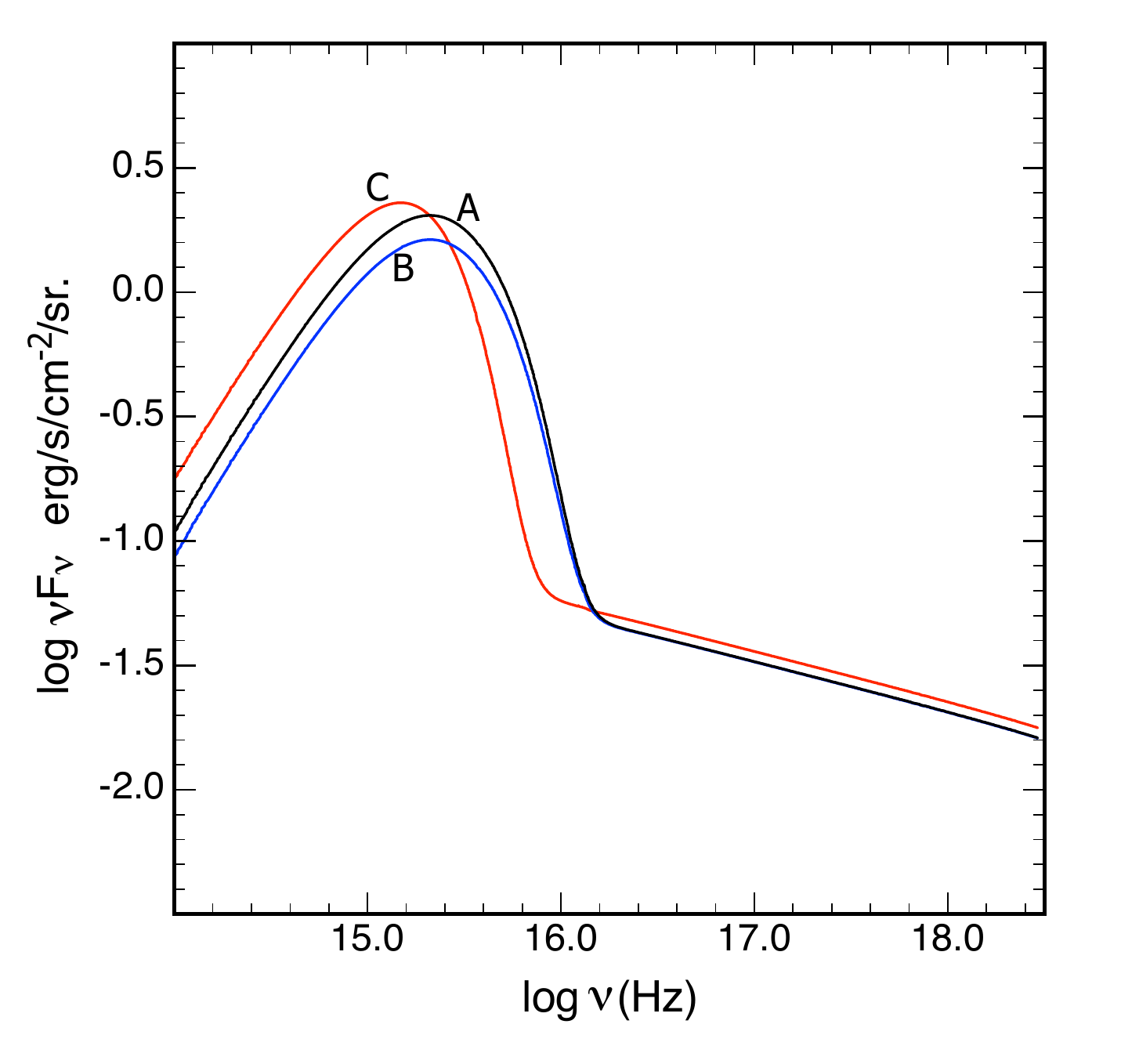}
  \caption{Inferred SED for the best-fit model of the nuclear spectrum of NGC~5427 (model A:  $M_{\rm BH} = 5\times 10^7 M_{\odot}$, an Eddington fraction of 0.1, a hard X-ray scale factor of 0.2 and $\log U =-1.3$.). The vertical scaling factor applies at the inner edge of the model NLR. Note the absence of an intermediate Compton component between $16 < \log \nu <18$. Such a component would arise naturally in Compton-heated optically-thin photoionised gas close to the ionising source. For comparison, we also show the other models which give a reasonable fit to the observed nuclear spectrum. Model B has the parameters $M_{\rm BH} = 5\times 10^7 M_{\odot}$, an Eddington fraction of 0.1 and a hard X-ray scale factor of 0.25 and $\log U =-1.5$, and model C has $M_{\rm BH} = 10^8 M_{\odot}$, an Eddington fraction of 0.05 and a hard X-ray scale factor of 0.2 and $\log U =-2.0$. The line fluxes for these are given in detail in Table \ref{table_2}.
 }
  \label{fig:SED}
  \end{figure}

 With these abundances, the best-fit model (Model A) has $M_{\rm BH} = 5\times 10^7 M_{\odot}$, an Eddington fraction of 0.1, a hard X-ray scale factor of 0.2, $\log U =-1.3$ and $\log{\rm P/k} = 6.6$. The corresponding source SED is shown in Figure \ref{fig:SED}. Reasonably satisfactory fits are also obtained for the parameters $M_{\rm BH} = 5\times 10^7 M_{\odot}$, an Eddington fraction of 0.1 and a hard X-ray scale factor of 0.25 and $\log U =-1.5$ (Model B), and $M_{\rm BH} = 10^8 M_{\odot}$, an Eddington fraction of 0.05 and a hard X-ray scale factor of 0.2 and $\log U =-2.0$ (Model C).
 
We also investigated the fits for the case of a $\kappa$-distribution of electrons rather than the assumption of a Maxwell-Boltzmann distribution. The empirical justification for adopting a non-thermal distribution in the case of \ion{H}{ii} regions has been discussed by \citet{Nicholls:2012apj,Dopita:2013apjs}. { Although we have no direct evidence that a $\kappa$-distribution applies to either \ion{H}{ii} regions or ENLR, such a distribution naturally arises wherever energy is injected locally  by long-range energy transport mechanisms such as cosmic rays, relativistic electrons, winds, magnetic waves or magnetic re-connection. The theoretical basis of these distributions has been shown to arise naturally from entropy considerations \citep{Tsallis95, Treumann99, Leubner02}, as well as in the  analysis by \citet{Livadiotis09}.  They explored the $q$ non-extensive statistical mechanics resulting from long-range energy transport mechanisms,  and have shown that $\kappa$ energy distributions arise as a consequence of this entropy formalism, in the same way as the Maxwell-Boltzmann distribution arises from Boltzmann-Gibbs statistics.}

 In our analysis we use $\kappa = 20$. For the fitting we must use a higher initial nuclear abundance; $3.75Z_{\odot}$, derived from the \ion{H}{ii} region abundance gradient data in Section (3.3) above. In this case, the best fit is obtained with the nuclear He abundance enhanced by +0.17dex and  N enhanced by +0.25 dex, very similar to the factors derived for the Maxwell-Boltzmann case. The best fit model (Model D in Table \ref{table_2}) has $M_{\rm BH} = 5\times 10^7 M_{\odot}$, an Eddington fraction of 0.1, a hard X-ray scale factor of 0.2, $\log U =-1.3$ and $\log{\rm P/k} = 6.6$. This model fits the observed spectrum even better than Model A.
 
 The comparison of these models with the observed nuclear spectrum is shown in Table \ref{table_2}. Compared with earlier attempts to model individual objects \emph{e.g.} \citet{Allen:99apj}, the fit is excellent. The large difference between the model and the observations in the case of  [\ion{N}{i}] is likely caused by a modelling error, as this line is notoriously difficult to predict, as this line arises from the extended partially-ionised zone close to the ionisation front which is heated by Auger electrons from the hard X-rays. The computed temperature in this zone is somewhat unreliable, and the charge-exchange rate with H which governs the fractional ionisation of N depends critically upon this temperature.

\begin{table*}
\centering
\caption{Extracted nuclear line fluxes for NGC~5427, and the corresponding best-fit photoionisation models. Model A has $M_{\rm BH} = 5\times 10^7 M_{\odot}$, an Eddington fraction of 0.1, a hard X-ray scale factor of 0.2 and $\log U =-1.3$,  Model B has the parameters $M_{\rm BH} = 5\times 10^7 M_{\odot}$, an Eddington fraction of 0.1 and a hard X-ray scale factor of 0.25 and $\log U =-1.5$, and Model C has $M_{\rm BH} = 10^8 M_{\odot}$, an Eddington fraction of 0.05 and a hard X-ray scale factor of 0.2 and $\log U =-2.0$.  Model D is appropriate for a $\kappa-$distribution with $\kappa=20$, and has $M_{\rm BH} = 5\times 10^7 M_{\odot}$, an Eddington fraction of 0.1, and a hard X-ray scale factor of 0.2, $\log U =-1.3$. \label{table_2}}
\begin{tabular}{lcccccc}
\hline
$\lambda$ ({\AA}) & Line ID & Flux &  Model A &  Model B &  Model C & Model D\\

 \hline
3727,9 & [\ion{O}{ii}]  & $2.24 \pm 0.36$ &1.92 &1.72 &1.90& 2.48 \\
3868 &[\ion{Ne}{iii}]  & $0.97 \pm 0.14$ & 0.40 &0.52 & 0.32 & 0.48 \\
3889 & H5 + \ion{He}{i} & $0.28 \pm 0.04$ & 0.26 & 0.31 & 0.26 & 0.29\\
3970 & H$\epsilon$ + [[\ion{Ne}{iii}]  & $0.31 \pm 0.09$ & 0.33 &0.33 &0.21& 0.30\\
4101 & H$\delta$ & $0.24 \pm 0.04$ & 0.26& 0.26 &0.26 &0.26 \\
4340 & H$\gamma$ & $0.47 \pm 0.06$ & 0.46& 0.46 &0.46 & 0.46 \\
4363 & [\ion{O}{iii}]  &  --  & 0.05& 0.08 & 0.03 & 0.07\\
4685 & \ion{He}{ii} & $0.17 \pm 0.03$ & 0.14 & 0.18 & 0.14 & 0.17\\
4861 & H$\beta$ & $1.00 \pm 0.09$ & 1.00 & 1.00 &1.00 &1.00 \\
4959 &[\ion{O}{iii}] & $2.27 \pm 0.19$ & 2.40 & 3.18 &1.81&2.60 \\
5007 & [\ion{O}{iii}] & $6.87 \pm 0.57$ & 6.94 & 9.18 & 5.24 & 7.51 \\
5198,200 & [\ion{N}{i}] & $0.27 \pm 0.05$ & 0.10 & 0.17 & 0.07& 0.05 \\  
5875 & \ion{He}{i} & $0.27 \pm 0.04$ & 0.27& 0.28 & 0.22 &0.27 \\
6300 & [\ion{O}{i}]& $0.37 \pm 0.03$ & 0.70 & 1.10& 0.53 & 0.56\\
6363 &[\ion{O}{i}] & $0.15 \pm 0.02$ & 0.23 & 0.35 & 0.17 & 0.18 \\
6548 & [\ion{N}{ii}] & $0.97 \pm 0.06$ & 1.16 & 1.28 &1.31& 1.00\\
6563 & H$\alpha$ & $3.10 \pm 0.19$ & 2.94 & 2.93 &2.95 & 2.98 \\
6584 &[\ion{N}{ii}]& $3.04 \pm 0.19$ & 3.44 & 3.77 & 3.87 & 2.96 \\
6717 & [\ion{S}{ii}] & $0.68 \pm 0.04$ & 0.93 & 1.33 & 0.82& 0.90\\
6731 & [[\ion{S}{ii}] & $0.63 \pm 0.04$ & 0.82 &1.21 & 0.76 & 0.82 \\
\hline
\end{tabular}
\end{table*}

\subsection{The luminosity of the black hole}
Our best fit model to the emission line spectrum inside a $3\times3$ arc sec. box surrounding the nucleus implies a bolometric luminosity of $\log L_{\mathrm bol.} = 44.3$ erg~s$^{-1}$. In this best-fit model, 25\% of the bolometric flux is in hydrogen ionising frequencies ($h\nu > 13.6$eV). Thus, the UV to far-IR luminosity is  $\log L_{\mathrm bol.} = 44.17$ erg~s$^{-1}$. This should be compared with the \citep{Woo:2002aa} value of $\log (L_\mathrm{bol.}) = 44.12\ \mathrm{erg\ s^{-1}}$ obtained by direct integration of the observed flux over this part of the SED. { In this context, it should be remembered that the galaxy is a Seyfert 2, so direct observation of the inner accretion disk and therefore of the big blue bump is impossible. The \citet{Woo:2002aa} integration relies upon the far-IR dust re-emission to derive an estimate of the bolometric  luminosity, and will therefore fail to include some proportion of the flux in the ionising continuum which escapes from the vicinity of the central engine.}

The inner empty zone in the model (required in order to provide the correct value of the ionisation parameter in the NLR gas) is $6\times10^{20}$cm (194pc or 1.1arc~sec at the distance of NGC~5427). This can be compared with the mean radius inside the  $3\times3$ arc~sec aperture (defined as the radius of a circle with total area 4.5 arc~sec$^2$) which is 1.2 arc~sec (211 pc). Clearly both the computed scale of the ENLR and the luminosity given by the model are consistent with the observations.

{ Given that we are in the radiation pressure dominated regime we can also estimate the luminosity from both the pressure in the ionised gas, $P$ and the distance of the ionised gas from the central engine, $r$, using 
\begin{equation}
P \sim nkT \sim L_{\mathrm bol.}/4\pi r^2c
\end{equation}
where $P$ is the gas pressure, $n$ is the total particle density, $k$ is the Boltzmann constant, and $c$ is the speed of light.  From the measured electron density of the [\ion{S}{ii}] gas, $n_{[S II]} = 330\pm 50$cm$^{-3}$, and using the theoretical model, } we determined the parameters of the gas in the [\ion{S}{ii}]  emitting zone, $n=1400$cm$^{-3}$, and $T=7450$K, giving a total pressure of $P=1.4\times10^{-9}$dynes cm$^{-2}$. Taking the radius as 194 - 211 pc, and allowing for the uncertainty in the density determination, we conclude that $\log L_{\mathrm bol.} = 44.3 \pm 0.1$ erg~s$^{-1}$ by this method - exactly the same as that previously derived. The fact that we can successfully estimate the luminosity from \emph{only} the [\ion{S}{ii}] density and the observed radius of the emitting gas from the central engine (provided that we are sure that the NLR is in the pressure-dominated regime) suggests that this method could find more general application to other Seyfert ENLR regions.

\subsection{The ENLR of NGC~5427}\label{sec:composite}
The extended narrow-line region (ENLR) of NGC~5427 is found out to 10\arcsec\ from the nucleus. It is immediately evident from Figure \ref{fig:BPT} and the left-hand panel of Figure \ref{fig:fit} that the position of the observed points on the BPT diagram cannot be simply explained by increasing dilution of the radiation field (decreasing $U$) as a function of distance from the AGN. However, other studies \citep{Scharwachter:2011aa,Davies:14mnras} have shown that the line ratios in the BPT diagrams can be understood as mixing between a NLR spectrum which is effectively invariant with radius, and a background contribution of  \ion{H}{ii} regions. In this section we investigate whether such a model also applies in the case of NGC~5427. 

We have computed model mixing curves based on the parameters obtained from the photoionisation models. For the \ion{H}{ii} regions, we use our inferred nuclear total oxygen abundance of $12 + \log(\rm O/H) = 9.24$ (corresponding to a metallicity of $Z = 3.7 Z_\odot$). We have computed the positions on the BPT diagrams for two ionisation parameters; $\log U = -3.5$ and -2.75  (corresponding to $\log q =7.0$ and 7.75). This range in ionisation parameter encompasses the measured values for all the \ion{H}{ii} regions listed in Table \ref{table_1}. (The actual observed range of ionisation parameters for these  \ion{H}{ii} regions is $\log U = -3.4$ to -3.1, corresponding to $\log q =7.1 - 7.4$). For the AGN, we use our best-fitting model A from Table~\ref{table_2}.

\begin{figure*}
   \centering
   \includegraphics[width=0.95\textwidth]{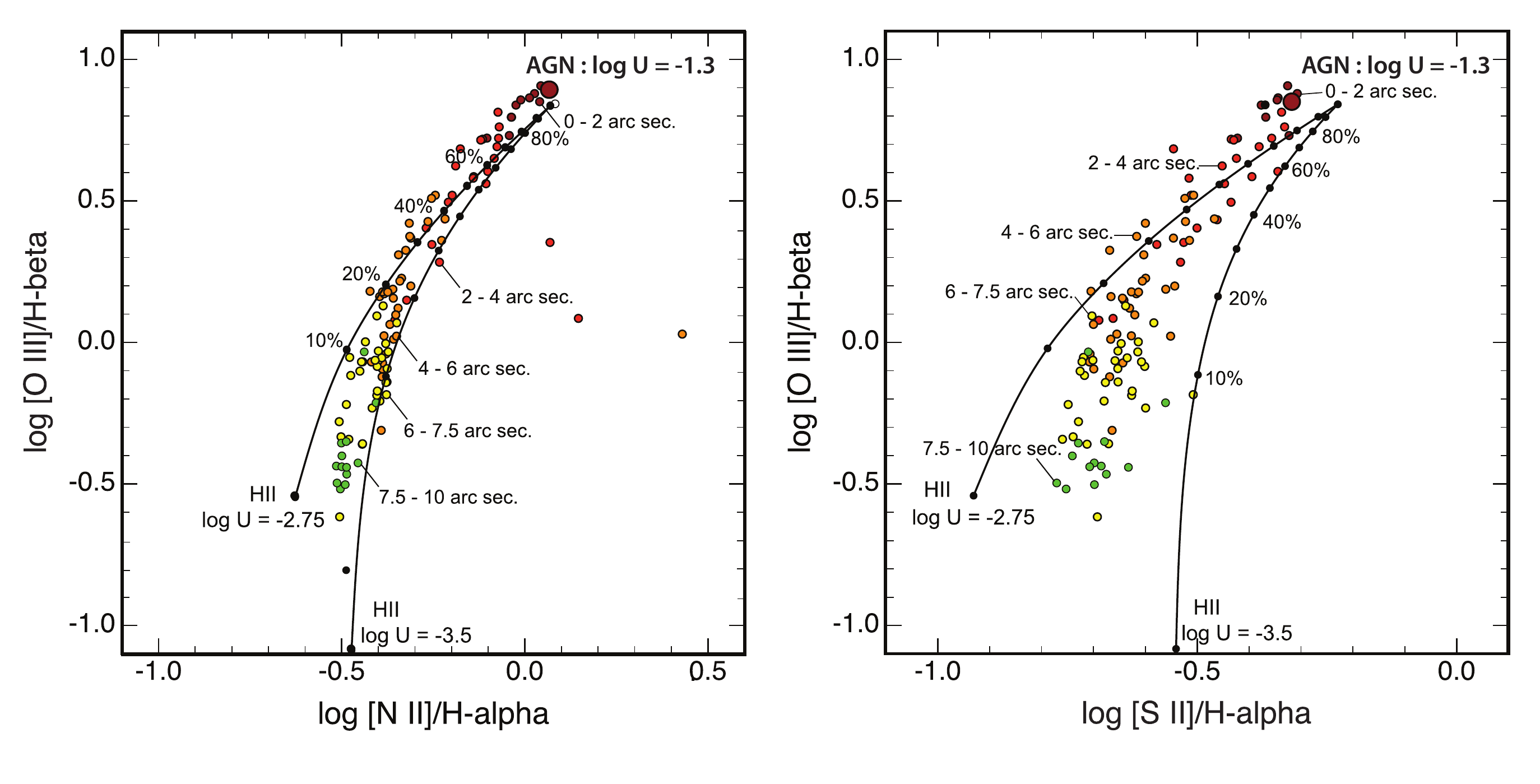}
  \caption{Representative mixing curves for all spaxels classified as "composite" based on their line ratios. The mixing curves are shown in the [\ion{N}{ii}]/H$\alpha$ versus [\ion{O}{iii}]/H$\beta$ diagnostic diagram. The observed emission-line ratios are colour-coded based on their de-projected distance from the nucleus. The distance ranges corresponding to the different colours are indicated in the plot. For clarity, only two representative mixing curves are shown. The anchor point for AGN ionisation is defined by $\log U = -1.1$ for the best-fitting model given in Table \ref{table_2}. The two anchor points for photoionisation in \ion{H}{ii} regions are defined by $\log U = -2.75$ and $\log U = -3.5$ for an oxygen abundance of $12+\log(\mathrm{O/H}) = 9.24$. }
  \label{fig:BPTmix}
  \end{figure*}

The line ratios predicted for mixtures of these two ionisation mechanisms are plotted in the classical BPT diagrams, [\ion{N}{ii}]/H$\alpha$ vs. [\ion{O}{iii}]/H$\beta$ and [\ion{S}{ii}]/H$\alpha$ vs. [\ion{O}{iii}]/H$\beta$ in Figure~\ref{fig:BPTmix}. It can be seen that the mixing curves provide an excellent fit to the observed shape of the ``composite/AGN" branch for the spaxels of the circum-nuclear region out to a radius of 10\arcsec\ (1.77~kpc). We find clear evidence for a decrease in the ENLR fraction as a function of radial distance. This direct correlation between distance from the nucleus and relative ionisation contribution is further support for the scenario that the branch of ``composite" line ratios is a result of the blending within an individual spaxel of  a component due to the ENLR and a component due to ionisation by stars. Similar mixing between starburst and AGN in composite galaxies has been adduced using infrared diagnostics by \citet{Hill:99aj} and  X-rays by \citet{Panessa:05apj}, and the observed BPT mixing curve is almost identical to those found by \citet{Scharwachter:2011aa} and \citet{Davies:14mnras} in other Seyfert galaxies.

The mixing lines in Figure~\ref{fig:BPTmix} implicitly assume that the ENLR has an invariant spectrum. While this is difficult to explain with classical models, with radiation pressure dominated models the reason is easier to understand. In those models, the pressure in the radiation field much exceeds the gas pressure in the ISM, and the pressure in the photo ionised gas which produces the main optical forbidden lines is set by the radiation pressure associated with the radiation field which has been absorbed up to that point in the photo ionised plasma. Therefore, since the radiation pressure decreases as $r^{-2}$ going out from the source, the pressure in, say, the [\ion{S}{ii}] - emitting gas will also drop off as  $r^{-2}$, locking  $\log U$ to a fixed value. This will continue up to the point where the radiation pressure becomes comparable to the natural pressure in the ISM, which effectively determines the radius of influence of the ENLR.

Given that most ENLR have broader velocity widths than \ion{H}{ii} regions in the galactic disk, the \ion{H}{ii} component may be separated from the ENLR components using high-resolution spectroscopy. This will be an important aspect of our {Siding Spring Southern Seyfert Spectroscopic Snapshot Survey} (S7) referred to in the introduction, and such data will enable us further investigate the physics of the NLR of Seyfert galaxies.
  
\section{Conclusions}

In this paper we have presented high-quality IFU observations of the Seyfert galaxy NGC~5427. From these we have been able to measure the abundance gradient accurately and so constrain the chemical abundances of the Seyfert 2 nucleus. Using these abundances, we have built a detailed photoionisation model for the nuclear region. We have identified the sensitivity of the various observed line ratios to the input parameters of the model; the intensity of the hard X-ray component relative to the accretion disk component of the EUV/X-ray SED. We found no evidence for an intermediate-energy Comptonised component in this object. 

We discovered that, in order to obtain an excellent fit to the measured emission-line ratios, the He abundance needed to be enhanced above that found from the \ion{H}{ii} regions by +0.16~dex, and the N abundance by +0.3~dex. This implies that massive fast-rotating stars are being formed within the accretion flow near the black hole and chemically polluting both the extended ($\sim 100$pc) outer disk structure and the ENLR. Such stars could potentially provide a source of viscosity in the disk, enhancing the accretion rate between 100 and 10~pc (approximately), and possibly providing a solution to the ``angular momentum problem''.

The nuclear luminosity is loosely constrained by the models;  $\log L_{\mathrm bol.} = 44.3\pm 0.1$ erg~s$^{-1}$, of which about 25\% is able to ionise hydrogen. The far-IR to UV flux estimated from the model, $\log L = 44.17 \pm 0.10$ erg~s$^{-1}$, agrees closely with what has been estimated by direct integration of the observed fluxes over these wavebands ($\log L = 44.12$ erg~s$^{-1}$, \citet{Woo:2002aa}). The best fit model has a Black Hole with a mass of $5\times10^7 M_{\odot}$ radiating at $\sim 0.1$ of its Eddington luminosity.

Finally, we have established that the spectra observed for individual spaxels in the ENLR of NGC~5427 out to a radius of $\sim 2$~kpc can be understood as a mixing between the ENLR and background star formation activity giving rise to normal \ion{H}{ii} region-like spectra. The ENLR is extended more in the NW-SE direction, roughly perpendicularly to the nuclear ring of star formation. In this sense it forms a classical ``ionisation cone".

The methodology we have established in this paper is generic and applicable to other Seyfert galaxies and their ENLR. It establishes the utility of using the \ion{H}{ii} regions to constrain the chemical abundances in the nucleus, and providing strict limits on the form of the ionising spectrum of the central engine.
 
\begin{acknowledgements}
J.S. acknowledges the European Research Council for the Advanced Grant Program Num 267399-Momentum. M.D. and L.K. acknowledge the support of the Australian Research Council (ARC) through Discovery project DP130103925.  This work was also funded in part by the Deanship of Scientific Research (DSR), King Abdulaziz University, under grant No. (5-130/1433 HiCi). M.D. and H.B. acknowledge this financial support.
\end{acknowledgements}

\bibliographystyle{aa} 
\bibliography{mybib.bib} 

\end{document}